\documentclass[12pt]{article}
\pdfoutput=1
\usepackage{xcolor}
\usepackage{hyperref}

\usepackage{setspace}
\usepackage{amsmath, amssymb, amsthm, float,graphicx}
\numberwithin{equation}{section}

\hypersetup{colorlinks=false, linkcolor=blue, citecolor=red}

\def\beq{\begin{eqnarray}}\def\eeq{\end{eqnarray}}
\def\be{\begin{equation}}\def\ee{\end{equation}}
\def\g{\gamma}
\def\r{\rho}
\def\s{\sigma}
\def\m{\mu}
\def\n{\nu}
\def\a{\alpha}
\def\e{\epsilon}
\def\k{\kappa}
\def\b{\beta}
\def\d{\delta}
\def\c{\chi}
\def\f{\phi}
\def\t{\theta}
\def\D{\Delta}
\def\l{\lambda}
\def\pd{\partial}

\def\la{\langle}
\def\ra{\rangle}
\def\fin{f_{\infty}}
\def\tr{{\rm tr~}}
\begin{document}
\title{\bf Holographic stress tensor at finite coupling }
\date{}

\author{Kallol Sen\footnote{kallol@cts.iisc.ernet.in} ~and Aninda Sinha\footnote{asinha@cts.iisc.ernet.in}\\ ~~~~\\
\it Centre for High Energy Physics,
\it Indian Institute of Science,\\ \it C.V. Raman Avenue, Bangalore 560012, India. \\}
\maketitle

\vskip 1cm
\begin{abstract}{\small  }
We calculate one, two and three point functions of the holographic stress tensor for any bulk Lagrangian of the form ${\mathcal{L}}(g^{ab}, R_{abcd}, \nabla_e R_{abcd})$. Using the first law of entanglement, a simple method has recently been proposed to compute the holographic stress tensor arising from a higher derivative gravity dual. The stress tensor is proportional to a dimension dependent factor which depends on the higher derivative couplings. In this paper, we identify this proportionality constant with a B-type trace anomaly in even dimensions for any bulk Lagrangian of the above form. This in turn relates to ${\mathcal{C}}_T$, the coefficient appearing in the two point function of stress tensors. We use a background field method to compute the two and three point function of stress tensors for any bulk Lagrangian of the above form in arbitrary dimensions. As an application 
we consider general situations where $\eta/s$ for holographic plasmas is less than the KSS bound.
\end{abstract}

\tableofcontents

\onehalfspacing
\section{Introduction}

Holographic methods have proved to be enormously useful to gain intuition about certain physical questions at strong coupling \cite{revs}. However, in most applications, attention has focused on cases where the bulk gravitational dual is two derivative. In the AdS/CFT dictionary, this corresponds to infinite 't Hooft coupling and an infinite number of colours. In order to make possible contact with the real world, one needs to consider effects due to finite 't Hooft coupling and a finite number of colours. In the best studied example corresponding to the ${\mathcal N}=4$ SU(N) supersymmetric Yang-Mills with the gravitational  dual being type IIB superstring theory on AdS$_5 \times $ S$^5$, finite coupling corrections correspond to specific higher derivative corrections in the low energy effective action of string theory. In addition to these corrections, there are also non-local contributions, for example from graviton loops. 
One essential step to take into account the contributions at finite coupling, is to be able to compute the holographic stress tensor and its correlation functions for an arbitrary higher derivative theory of gravity.

Calculating the holographic stress tensor itself from first principles \cite{skenderis} appears to be prohibitively difficult except for certain cases where the generalized Gibbons-Hawking term and the counterterms are known. Since the generalized Gibbons-Hawking term is not known for an arbitrary higher curvature theory, this stymies any progress using conventional approaches--only some sporadic results for specific examples are known in the literature \cite{HT, others}. Recently, a way around has been found using the first law of entanglement pertaining to spherical entangling surfaces \cite{FM}. 
The way this works is as follows: The first law of entanglement states that \cite{tadashi}
\be\label{1stlaw}
\Delta S= \Delta H\,,
\ee
where for two density matrices $\rho, \sigma$ with $\sigma\equiv e^{-H}/\tr e^{-H}$ being the reduced density matrix for a spherical entangling surface in a CFT with $H$ being the modular hamiltonian, $\Delta H=\langle H\rangle_\rho -\langle H\rangle_\sigma$ and $\Delta S= S(\rho)-S(\sigma)$ with $S(\rho)=-\tr \rho \log \rho$ being the von Neumann entropy for $\rho$ and is the entanglement entropy for a reduced density matrix $\rho$. The equality arises at linear order in perturbation, meaning that $\rho,\sigma$ belong to some family of density matrices $\hat\rho$ parametrized by some perturbation parameter $\lambda$ such that $\sigma=\hat\rho(\lambda=0), \rho=\hat\rho(\lambda)$ and we are interested only in linear order in $\lambda$. At nonlinear order in $\lambda$ we get an inequality which corresponds to the positivity of relative entropy, leading to $\Delta H>\Delta S$. The expression for $H$ (which will be given below) involves the time-time component of the field theory stress tensor. In holography, for spherical entangling surface, the entanglement entropy for the vacuum state across the sphere $S^{d-2}$ gets mapped to the thermal entropy on $R\times H^{d-1}$. Using the gravitational dual, the thermal entropy is computed using the Wald entropy which is known for an arbitrary higher derivative theory of gravity. For bifurcate Killing horizons, there is a theorem due to Iyer and Wald \cite{IW} which states that linearized perturbations satisfy the first law of thermodynamics which translated to our case means that eq.(\ref{1stlaw}) would be applicable with linearized perturbations in the Wald formula. Thus the LHS of eq.(\ref{1stlaw}) can be computed using the linearization of the Wald formula. The RHS of eq.(\ref{1stlaw}) has the perturbation of the time-time component of the field theory stress tensor which now can be determined. In order to be able to do the integral, one approximates the entangling surface radius $R$ to be small. Since the only dimensionless parameter is $R^d \langle T_{\mu\nu}\rangle$, the perturbative expansion can be done either by treating $R$ to be small or by treating $\langle T_{\mu\nu}\rangle$ to be small. Thus although the expression for the stress tensor obtained using the above logic pertains to an excited state that is a small perturbation from the vacuum state, the expression should hold for any $\langle T_{\mu\nu}\rangle$. Since this can be done for an arbitrary higher derivative theory, we thus know how to compute the holographic stress tensor for such a bulk dual.

The result of this calculation is a very compact expression for the stress tensor in terms of certain parameters appearing in the linearized expression of the Wald formula. In particular, if one ignores covariant derivatives of the curvature tensor, the result can be worked out quite simply. Writing the linearized Wald functional as 
\be\label{eq2}
\d E^{abcd}_R=-c_2 g^{\la ab}g^{cd\ra}h-c_3 h^{\la ab}g^{cd\ra}+ c_4 g^{\la ab}g^{cd\ra}{\mathcal{R}}+ c_5 {\mathcal{R}}^{\la ab}g^{cd\ra}+ c_6 {\mathcal{R}}^{abcd},
\ee
where $c_2$ and $c_3$ are not independent coefficients but related to the other coefficients as,
\be 
c_2=-2d c_4-c_5,\ \ c_3=2c_1-(d-1)c_5-4c_6\,,
\ee
which we will demonstate later, $c_1$ is defined through the Wald function $E_R^{abcd}=c_1g^{\la ab}g^{cd\ra}$ and where,
\be 
\d g^{ab}=-h^{ab}, \ \ \d R^{abcd}={\mathcal{R}}^{abcd}, \ \ h=g_{cd}h^{cd},
\ee
one finds that\footnote{ Note that we are considering field theory in flat space.}\footnote{ The normalization of $\D H$ is fixed by the definition of modular Hamiltonian in \eqref{1stlaw}. On the RHS the normalization of $\D S$ is fixed by holography where we demand that the definition of $S$ gives the correct universal terms. This resolves any ambiguity in the definition of $\la T_{\m\n}\ra$ in \eqref{eq1} by using $h_{\m\n}^{(d)}$ in \eqref{FG}.}
\be \label{eq1}
\la T_{\m\n} \ra=d \tilde{L}^{d-3}[c_1+2(d-2)c_6]h^{(d)}_{\m\n},
\ee
where $h^{(d)}_{\m\n}$ appears in the Fefferman-Graham expansion of the asymptotic AdS metric as
\be\label{FG}
ds^2=\tilde L^2\frac{dz^2}{z^2}+\frac{1}{z^2}(g^{(0)}_{\mu\nu}+z^2 g^{(2)}_{\mu\nu}+\cdots z^d h^{(d)}_{\mu\nu}+\cdots)dx^\mu dx^\nu\,.
\ee

$\tilde{L}$ is the AdS radius. This begs the question: What is this simple proportionality constant depending on $c_i$'s in \eqref{eq1}? Since the linearized Wald functional was involved in the derivation of this simple form, with hindsight we can anticipate that there are simplifications waiting to happen if we consider rewriting the Lagrangian as a background field expansion around a suitable background. Recently this background field method has been used to find simple expressions for trace anomalies in even dimensions \cite{chinese}. We will make a simple modification to this method so that the anomaly calculation can be carried out easily using Mathematica. Let us now explain why this method is useful in correlating with the results above as well as calculating higher point correlation functions. Given a Lagrangian $ {\mathcal{L}}(g^{ab},R_{bcde}, \nabla_f R_{bcde},\cdots)$, we are going to treat $g_{ab}$ and $R_{abcd}$ as independent variables. We are going to expand $R_{abcd}$ around $\bar{R}_{abcd}=-\frac{1}{\tilde L^2}(g_{ac}g_{bd}-g_{ad}g_{bc})$ where $g_{ab}$ in this expression is the full metric. Raising and lowering indices and the covariant derivative is done using the full metric. Define $\D R_{abcd}=R_{abcd}-\bar{R}_{abcd}$. Then on the AdS background $(g^{(0)}_{\m\n}=\eta_{\m\n})$ this quantity is zero. Further if we linearize this then in the transverse traceless gauge, it can be easily checked that $(\D R_{ab})^L=(\D R)^L=0$. This is the reason why the expressions we will compute for the stress tensor correlation functions will take on simple forms.
Let us start with  $ {\mathcal{L}}(g^{ab},R_{bcde})$, i.e., no covariant derivatives (we will set the AdS radius $\tilde L=1$ from hereon and will reinstate it when needed). The Lagrangian after doing the background field expansion takes the form
\be\label{lag}
{\mathcal{L}}=(c_0+c_1\D R+\frac{c_4}{2} \D R^2+\frac{c_5}{2} \D R^{ab} \D R_{ab}+\frac{c_6}{2} \D R^{abcd}\D R_{abcd}+\sum_{i=1}^{8}\tilde{c}_i\D {\mathcal{K}}_i+\cdots),
\ee
where $c_0=-2d c_1$ from lowest order equations of motion and $\D {\mathcal{K}}_{i}={\mathcal{K}}_{i}|_{R\rightarrow(R-\bar{R})}$. Note that we are \textit{not} treating $c_i$'s perturbatively.  This Lagrangian can be shown to lead to \eqref{eq2}. The basis for the third order terms is given by 
\beq\label{Ki}
{\mathcal{K}}_i=(R^3,R^a_b R^b_c R^c_a, R R^{ab}R_{ab}, RR^{abcd}R_{abcd}, R^{ab}R^{cd}R_{acbd}, R_{ab}R^{acde}R^{b}_{ \ cde},\nonumber\\ R_{abcd}R^{abef}R^{cd}_{ \ \ ef}, R_{abcd}R^{aefc}R^{b \ \ d}_{ \ ef} ).
\eeq
We are not using an explicit overall factor of $1/2\ell_p^{d-1}$ with the action since all the coefficients in the action are assumed to implicitly have the factor. In order to compute $n$-point functions we expand the bulk action up to $n$'th order in the perturbation. However, since $\D R_{abcd}^{AdS}=0$, this means that we only need to retain up to $O((\D R)^n)$ terms in the Lagrangian. Thus, the background field expanded Lagrangian is an expansion in terms of the correlation functions of the stress tensor. Further simplifications happen. Consider $(\Delta R)^2$ or $(\Delta R_{ab})^2$. Since the linearized $\D R$ and $\D R_{ab}$ both vanish, these terms can only contribute to four-point functions onwards. Thus we do not expect $c_4$ or $c_5$ in eq.(\ref{lag}) above to enter the one, two or three point functions. This is consistent with the absence of these coefficients in eq.(\ref{eq1}). Moreover, this conclusion does not change on including covariant derivatives.

Let us now summarize our findings for the correlation functions that follow from the Lagrangian in eq.(\ref{lag}). 
The stress tensor two point function takes the form 
\be
\la T_{ab}(x)T_{cd}(x')\ra=\frac{{\mathcal{C}}_T}{|x-x'|^{2d}}{\mathcal{I}}_{ab,cd}(x-x'),
\ee
where 
\be
{\mathcal{I}}_{ab,cd}(x)=\frac{1}{2}[I_{ab}(x)I_{cd}(x)+I_{ad}(x)I_{bc}(x)]-\frac{1}{d}\eta_{ab}\eta_{cd},
\ee
and
\be
I_{ab}(x)=\eta_{ab}-2\frac{x_ax_b}{x^2}.
\ee
The coefficient ${\mathcal{C}}_T$ from the $d+1$ dimensional bulk Lagrangian works out to be
\be
{\mathcal{C}}_T= f_d\tilde{L}^{d-1}[c_1+2(d-2)c_6],
\ee
where $\tilde{L}$ is the AdS radius and $f_d$ is constant $d$ dependent factor given by \cite{holGB}
\be
f_d=2\frac{d+1}{d-1}\frac{\Gamma[d+1]}{\pi^{d/2}\Gamma[d/2]}.
\ee 
Thus the holographic stress tensor in eq.\eqref{eq1} can be written as 
\be
\la  T_{\m\n}\ra= \frac{d}{\tilde{L}^2f_d}{\mathcal{C}}_T h^{(d)}_{\m\n}.
\ee
Further for even dimensional CFTs the coefficient ${\mathcal{C}}_T$ is related to a B-type anomaly coefficient as we will show. In particular, the A-type Euler anomaly coefficient is simply proportional to $c_1$ while the B-type anomaly coefficient (conventionally called $c$ in 4$d$ and $B_3$ in 6$d$) is proportional to $C_T$.

We use the method of background field expansion to calculate the three point functions of stress tensor. Following the simple method devised in \cite{HM, hofman} and used in \cite{quasitop} we perform the calculation of the three point function in a shockwave background and obtain information about the three point function from the energy flux given by (these results are for $d\geq 4$, for $d=3$, the term proportional to $t_2$ is absent),

\be
\la \epsilon(\textbf{n})\ra=\frac{E}{4\pi \Omega_{d-2}}[1+t_2(\frac{\e_{ij}^* \e_{ik}n^jn^k}{\e_{ij}^* \e_{ij}}-\frac{1}{d-1})+t_4 (\frac{|\e_{ij}n^jn^k|^2}{\e_{ij}^*\e_{ij}}-\frac{2}{d^2-1})]\,,
\ee
where $\Omega_{d-2}$ is the volume of a unit $(d-1)$ sphere and,
\begin{align}
\begin{split}
t_2 &=\frac{d(d - 1)}{c_1+2(d-2)c_6} [2c_6-12 (3d + 4) \tilde c_7+3 (7 d + 4) \tilde c_8],\\
t_4&= \frac{6d(d^2 - 1)(d+2)}{c_1+2(d-2)c_6}(2\tilde c_7 -\tilde c_8).
\end{split}
\end{align}
$\textbf{n}$ is the unit normal in the direction in which energy flux is measured and $t_2 \ \text{and} \ t_4$ are determined holographically. The coefficients $t_2$, $t_4$ and $ {\mathcal{C}}_T$ are related to the three independent coefficients appearing in the three point functions \cite{osborn,erdmenger}
\footnote{The relation between $t_2 \ t_4  \ \text{and} \ {\mathcal{C}}_T$ and the CFT coefficients ${\mathcal{A}}, \ {\mathcal{B}} \ \text{and} \ {\mathcal{C}}$ are 
\begin{align}
\begin{split}
{\mathcal{C}}_T&=\frac{\Omega}{2d(d+2)}[(d-1)(d+2){\mathcal{A}}-2{\mathcal{B}}-4(d+1){\mathcal{C}}], \ t_2=\frac{2(d+1)}{d}\frac{(d-2)(d+2)(d+1){\mathcal{A}}+3d^2{\mathcal{B}}-4d(2d+1){\mathcal{C}}}{(d-1)(d+2){\mathcal{A}}-2{\mathcal{B}}-4(d+1){\mathcal{C}}}\\
t_4&=-\frac{d+1}{d}\frac{(d+2)(2d^2-3d-3){\mathcal{A}}+2d^2(d+2){\mathcal{B}}-4d(d+1)(d+2){\mathcal{C}}}{(d-1)(d+2){\mathcal{A}}-2{\mathcal{B}}-4(d+1){\mathcal{C}}},
\end{split}
\end{align}
}. Notice that for $d=4$, the $\tilde c_7, \tilde c_8$ dependence in $t_2$ and $t_4$ are packaged in the same way, namely as $2\tilde c_7-\tilde c_8$. This is indicative of the fact that $c,a,t_2, t_4$ satisfy the relation $(c-a)/c=t_2/6+4 t_4/45$. This relation enables one to extract the 4d Euler anomaly $a$ from the knowledge of two and three point functions. In six (and higher) dimensions, there is no such relation (in fact not even for a linear combination of the A-anomaly and the B-anomaly coefficients) indicating the fact that a similar relation involving the Euler anomaly coefficient will also involve higher point correlation functions.

We can easily extend the above results to the $ {\mathcal{L}}(g^{ab},R_{bcde}, \nabla_f R_{bcde})$ case, i.e., to the situation where there are at most two covariant derivatives of the curvature tensor in the action. 
First notice that since the linearized $\D R_{ab}$ and $\D R$ both vanish, only terms like $\nabla_e \D R_{abcd} \nabla^e \D R^{abcd}$ will contribute to the two and three point functions while only terms like $\D R_{....}\nabla \D R_{....}\nabla \D R_{....}$ will contribute to the three point functions. Further, we will show that using the Bianchi identities and integration by parts \cite{RM}, the$\nabla_e \D R_{abcd} \nabla^e \D R^{abcd}$ terms can be rewritten in terms of $(\Delta R_{....})^3$ and $\nabla_a \D R_{bc} \nabla^a \D R^{bc}$, $\nabla_a \D R \nabla^a \D R$. Since the last two terms do not contribute to two or three point functions, the result for the two point functions will involve a redefined $c_6$. We will explicitly show that the result for the three point functions also follows a similar simple trend.

As an application for our methods we will compute the ratio of shear viscosity ($\eta$) to entropy density ($s$) for a general four derivative bulk dual, without assuming the coupling constants to be small (for earlier related work see \cite{etabs}). Then following \cite{quasitop}, we will demand that $-3\leq t_2 \leq 3$ as well as ${\mathcal{C}}_T>0$, $s>0$. These constraints were sufficient in the Gauss-Bonnet case \cite{HM,vis} to lead to $\eta/s \geq \frac{16}{25}\frac{1}{4\pi}\approx 0.64 \frac{1}{4\pi}$. We will find that in the general four derivative case, we can tune the couplings so that these conditions are satisfied but $\eta/s$ is arbitrarily small. This is of course due to the fact that the underlying theory has non-unitary modes. We will also show that for the Weyl-squared theory, the above constraints lead to $\eta/s \geq \frac{12-3\sqrt{2}}{14}\frac{1}{4\pi}\approx 0.55 \frac{1}{4\pi}$ while including Weyl-cubed terms, the same constraints lead to $\eta/s\geq 0.17\frac{1}{4\pi}$. Both these theories will have non-unitary modes supported near the horizon. It is interesting to note that there is still a bound on the ratio in such theories.

The rest of the paper is organised as follows. In section\eqref{2} we give a brief review of the calculations of \cite{FM}. In section\eqref{3} we calculate the holographic trace anomalies by considering the background field expanded Lagrangian and also how various coefficients of the Lagrangian in \cite{FM} are related to the Lagrangian we are considering. We then compute the trace anomalies in $d=2,4,6$ and show that the B-type anomalies are the coefficients in the expression for the holographic stress tensor. In section\eqref{4} we extend the analysis to Lagrangians containing covariant derivatives on the Riemann tensors and show how the anomaly coefficients get modified. More specifically we show that $c_6$ in the B-type anomaly coefficients can be replaced by an effective $c'_6$ in presence of the $\nabla R$ terms in the Lagrangian. In section \eqref{5} and section\eqref{6} we extend the analysis to calculating the holographic two and three point functions of the stress tensor. We show that the coefficient in the holographic one point function of the stress tensor is related to the coefficient of the holographic two point functions of the stress tensor for arbitrary dimensions.  Section \eqref{ebs} presents one application of the method of background field expansion in the calculation $\eta/s$. We present the calculations for Weyl-squared, Weyl cubed and general $R^2$ gravity (appendix \eqref{etas}). We also show that the bounds for $\eta/s$ for these theories pertaining to the physical constraints satisfied by the three point functions are much smaller that the KSS bound \cite{KSS}. We end with a discussion about open problems in section\eqref{7}.  

\section{Stress tensor from first law of entanglement} 
In this section we review the derivation of the holographic stress tensor from the first law of entanglement \cite{tadashi} for \eqref{lag} following \cite{FM}.
The modular hamiltonian for a spherical entangling region of radius $R$ and centered around $x=x_0$, is given by
\be
H_A=2\pi\int_{A(R,x_0)} d^{d-1} x \frac{R^2-|\bf{x}-\bf{x_0}|^2}{2R}\la T_{tt}\ra,
\ee 
and for any perturbation around the CFT vacuum we have 
\be
\D H_A=2\pi\int_{A(R,x_0)} d^{d-1} x \frac{R^2-|\bf{x}-\bf{x_0}|^2}{2R}\d T_{tt}\,.
\ee 
\par
As mentioned in \cite{FM}, the entanglement entropy of the spherical entangling region in the vacuum CFT is equal to the entropy of a thermal CFT on a hyperbolic cylinder with the temperature set by the length scale of the hyperbolic spacetime. From the holographic side the thermal entropy is given by the entropy of the hyperbolic black hole, which for any classical higher derivative theory of gravity is evaluated using the Wald formula \cite{IW}
\be
S^{Wald}= -2\pi \int_{\mathcal{H}} d^n\s \sqrt{h}\frac{\d{\mathcal{L}}}{\d R^{ab}_{cd}} n^{ab}n_{cd},
\ee
where ${\mathcal{L}}$ is given in \eqref{lag} and $n^{ab}$ is the unit binormal to the bifurcate Killing horizon ${\mathcal{H}}$. 
In general the Wald entropy functional differs from the enanglement entropy functional by squares of the extrinsic curvature \cite{sinha} but for the spherical entangling region these terms vanish and $S^{Wald}=S_{EE}$ at the linear order in perturbations \cite{FM}. Further, the perturbations of the vacuum CFT imply perturbations of the thermal CFT since the perturbations of the vacuum AdS imply perturbations of each of the AdS-Rindler wedges for the thermal state\footnote{ This assumption is only valid at the leading order in perturbations. In the next order the hyperbolic horizon changes due to the perturbations and we do not have the AdS-Rindler patch}. 
\par
Before proceeding we will specify the notations and conventions. Throughout we set the AdS radius\footnote{Note that $L$ is the length associated with the cosmological constant}.$\tilde{L}=1$ except where we explicitly restore it on dimensional grounds. $R$ is the radius of the entangling ball.  
In terms of the Poincar$\acute{e}$ coordinates, AdS spacetime is given by,
\be
ds^2=\frac{dz^2+\eta_{\m\n}dx^{\m}dx^{\n}}{z^2}.
\ee
The spherical entangling region $A$ in the vacuum CFT is associated with the hemispherical region $\tilde{A}$ in the black hole background given by
$\tilde{A}=\{t=t_0, \ (x^i-x_0^i)^2+z^2=R^2\}$ in Poincar$\acute{e}$ coordinates. These two different regions have the same boundary $\pd A$ in the boundary CFT. Thus $S_{EE}$ is equal to $S^{Wald}$ evaluated on $\tilde{A}$. Similarly the perturbation $\D S_{EE}$ of the vacuum CFT is equal to $\d S^{wald}$. For holographic CFTs the gravitational version of $\d S_A=\d E_A$ is given by $\d S^{grav}=\d E^{grav}=\d S^{Wald}$ and can be used to relate the $\la T_{\m\n}\ra$ to the asymptotic form of the metric in the holographic side. In the limit of $R\rightarrow 0$, $\d \la T_{tt}(t_0,\bf{x})\ra$ can be replaced by its central value $\d \la T_{tt}(x_0)\ra$ and we have using $\d E_{A}=\d S_A$,
\be\label{Ttt}
\d\la T_{tt}(x_0)\ra=\frac{d^2-1}{2\pi \Omega_{d-2}}\lim_{R\rightarrow0}(\frac{1}{R^d}\d S_A^{Wald}),
\ee
and repeating for arbitrary Lorentz frames we have
\be
u^{\m}u^{\n}\d\la T_{\m\n}(x_0)\ra=\frac{d^2-1}{2\pi \Omega_{d-2}}\lim_{R\rightarrow0}(\frac{1}{R^d}\d S_A^{Wald}).
\ee

The variation of the Wald entropy around the hyperbolic black hole background for an arbitrary higher derivative theory of gravity is given by
\be \label{swald}
\d S^{Wald}=\d(-2\pi\int_{\tilde{A}}E^{abcd}_R \epsilon_{ab}n_{cd}),
\ee
where $E^{abcd}$ is the Wald functional of the curvatures and their covariant derivatives, $\epsilon_{ab}$ is the volume element and $n_{cd}=n^1_cn^2_d-n^1_dn^2_c$ is the binormal to the bifurcation surface $\bar{B}$ respectively.

\subsection{For ${\mathcal{L}}(g^{a b},R_{cdef})$}\label{2}
 Evaluated on an AdS background where $R_{abcd}=-(g_{ac}g_{bd}-g_{ad}g_{bc})$, it can be shown (see appendix \ref{A1}) that the Wald functional and its linear variation for \eqref{lag} (without covariant derivatives of curvature terms in the action) takes the simple form 
\be\label{ewald}
E^{abcd}_{R}=c_1 g^{\la ab}g^{cd\ra}\,,
\ee
and,
\be
\d E^{abcd}_R=-c_2 g^{\la ab}g^{cd\ra}h-c_3 h^{\la ab}g^{cd\ra}+ c_4 g^{\la ab}g^{cd\ra}{\mathcal{R}}+ c_5 {\mathcal{R}}^{\la ab}g^{cd\ra}+ c_6 {\mathcal{R}}^{abcd},
\ee
where all the coefficients are not independent but related by \cite{FM}
\be 
c_2=-2d c_4-c_5,\ \ c_3=2c_1-(d-1)c_5-4c_6\,,
\ee
and $\la,\ra$ implies that it has been properly (anti)symmetrized to have the properties of the Riemann tensor. The linearized Reimann tensor is given by
\be\label{linriemann}
{\mathcal{R}}_{abcd}=\frac{1}{2}(\nabla_c\nabla_b h_{ad}-\nabla_d\nabla_b h_{ac}+\nabla_d\nabla_a h_{bc}-\nabla_c\nabla_a h_{bd})+\frac{1}{2}(R_{aecd}h^{e}_{b}+R^e_{bcd}h_{ae}).
\ee
When $\d T_{\m\n}$ is small and $R\rightarrow0$, the scaling analysis in \cite{FM} shows that at the leading order we can neglect all the derivatives $\pd_{\m\neq z}$ in comparison to $\pd_z$. Near the boundary, the metric perturbations can be written as
\be
h_{\m\n}(z,x^\l)=z^{\D-2}h_{\m\n}(x^{\l})+\dots
\ee
Using \eqref{linriemann} the relevant components of the linearized Wald functional in\eqref{eq2} take the form
\be\label{ew}
\d E_R^{(1)\m z\n z}=A h^{\m\n}g^{zz}+B h g^{\m\n}g^{zz}, \ \d E_R^{(1)\m\n\r\s}=C h g^{\la \m\n}g^{\r\s\ra}+D h^{\la \m\n}g^{\r\s\ra}\,,
\ee
where the coefficients $A,B,C,D$ are functions of the coefficients\footnote{ The explicit dependences of $A,B,C \ \text{and} \ D$ on the coefficients $c_1\dots c_6$ are given in footnote 20 of \cite{FM}} in \eqref{lag}. Substituting \eqref{ew} and \eqref{ewald} into \eqref{swald}, we get,
\be\label{intwald}
\d S^{Wald}=\frac{4\pi \tilde{L}^{d-3}}{R}\int_{A}\frac{d^{d-1}x}{z^{d-2}}(A_1 h_{tt}+A_2 \eta^{\m\n}h_{\m\n})\,.
\ee
After putting $\D=d$ in order to get a finite answer \cite{FM} we find
\begin{align}
\begin{split}
A_1&= 2(A-\frac{D}{4})(d-2)R^2+[\frac{c_1}{2}-\frac{D}{2}(d-2)+2A(d-1)](\frac{|x|^2}{d-1}-R^2),\\
A_2&=(\frac{c_1}{2}+2B)R^2+[\frac{c_1}{2}+\frac{D}{2}+(C-2B)(d-1)]\frac{|x|^2}{d-1}.
\end{split}
\end{align}
Performing the integral in \eqref{intwald} and using \eqref{Ttt} we have, 
\be
\d T^{grav}_{tt}=\alpha h^{(d)}_{tt}+\beta \eta_{tt}h^{(d)\m}_{\m},
\ee
where the coefficients are given as
\begin{align}
\begin{split}
\a&=d(-c_1+c_3+(d-1)c_5+2d c_6)\,,\\
\b&=[-(d+2)c_1+2(d+1)c_2+c_3+2d(d+1)c_4+(d+1)c_5-2(d-2)c_6]\,,
\end{split}
\end{align}
and $h^{(d)}_{\mu\nu}$ has no $z$ dependence. These can be generalized to an arbitrary Lorentz frame and combined with the tracelessness and conservation equations $h^{(d)\m}_{\m}=0, \ \pd_{\m}h^{(d)\m\n}=0$ we have \eqref{eq1} as
\be
\d T^{grav}_{\m\n}=d\tilde{L}^{d-3}[c_1+2(d-2)c_6]h^{(d)}_{\m\n}.
\ee

\subsection{For ${\mathcal{L}}(g^{ab}, R_{cdef},\nabla_a R_{bcde})$}\label{nabla}

The above analysis can be extended to actions containing covariant derivatives on the Riemann tensors. The most general term containing arbitrary covariant derivatives on the curvature tensors is deferred for futute work; we consider here ${\mathcal{L}}(g^{ab}, R_{cdef},\nabla_g R_{cdef})$. The background field expansion of the action at $O((\Delta R)^2)$ is given by
\be
S_{\nabla R}=\int d^{d+1}x\sqrt{g}Z^{efabcdmnrs}\nabla_e \D R_{abcd}\nabla_f\D R_{mnrs}.
\ee
It can be shown (see appendix B) that the above action can be written as
\be
S=\int d^{d+1}x\sqrt{g}[d_1 \D R^{abcd}\nabla^2\D R_{abcd}+d_2\D R^{ab}\nabla^2 \D R_{ab}+d_3\D R\nabla^2\D R].
\ee
Since $\nabla_a g_{bc}=0$, we can write
\be\label{nablar}
S=\int d^{d+1}x\sqrt{g}[d_1 \D R^{abcd}\nabla^2R_{abcd}+d_2\D R^{ab}\nabla^2R_{ab}+d_3 \D R\nabla^2R].
\ee
At the linear order in fluctuations using \eqref{linriemann}
\begin{align}
\begin{split}
{\mathcal{R}}^L_{ab}&=\frac{1}{2}\nabla^c\nabla_{a}h_{bc}+\frac{1}{2}\nabla_b\nabla^dh_{ad}-\frac{1}{2}\nabla^2 h_{ab}-\frac{1}{2}\nabla_a\nabla_b h+\frac{1}{2}(R_{aebd}h^{ed}+R^e_bh_{ae}),\\
{\mathcal{R}}^L&=\nabla^a\nabla^bh_{ab}-\nabla^2h-d h.
\end{split}
\end{align}
If we consider the transverse, traceless gauge $\nabla^a h_{ab}=0, h_a^a=0$, we have,
\be
{\mathcal{R}}^L=0, \ \text{and} \ {\mathcal{R}}^L_{ab}=-[\frac{1}{2}\Box+d]h_{ab}.
\ee
We can see that $\nabla^2 R$ term will not contribute to the action. To see that $(\nabla_a \D R_{bc})^2$ will also not contribute to the action, we will first carry out the linearization of $\D R_{ab}$ which is given by
\be
\D R_{abcd}^L={\mathcal{R}}^{L}_{abcd}-\bar{R}^L_{abcd}={\mathcal{R}}^L_{abcd}+(g^{(0)}_{ac}h_{bd}+g^{(0)}_{bd}h_{ac}-g^{(0)}_{ad}h_{bc}-g^{(0)}_{bc}h_{ad})\,.
\ee
Contracting with $g^{(0)bd}$ we get,
\be
\D R^L_{ac}={\mathcal{R}}^L_{ac}+(d-1)h_{ac}=-\frac{1}{2}[\Box+2]h_{ac}\,.
\ee
This term vanishes on using the lowest order equation of motion for $h_{ab}$. Thus this term does not contribute to the holographic stress tensor. For the remaining $\nabla_a R_{bcde}$ terms we can use the Bianchi Identity as in \cite{RM} to put the final expression [see appendix \eqref{bianchi}] in the form (neglecting the total derivatives),
\beq
R^{bcde}\nabla^2R_{bcde}=-4(\nabla_dR_{ce})^2+(\nabla R)^2-4R^{cd}R^{e \ \ f}_{\ dc \ }R_{ef}-4R^c_dR^d_fR^f_c+2R^{bcde}R^f_bR_{fcde}\nonumber\\+2R^{bcde}R^{a \ \ f}_{ \ bc \ }R_{afde}+4R^{bcde}R^{a \ \ f}_{ \ bd \ }R_{acfe}.
\eeq
The $O(R^3)$ terms in the expression are given by
\be
S_{R^3}=-4R^{cd}R^{e \ \ f}_{\ dc \ }R_{ef}-4R^c_dR^d_fR^f_c+2R^{bcde}R^f_bR_{fcde}+2R^{bcde}R^{a \ \ f}_{ \ bc \ }R_{afde}+4R^{bcde}R^{a \ \ f}_{ \ bd \ }R_{acfe}.
\ee
Doing a similar backgroound field expansion of the above terms we have at the second order in the expansion,
\be
O(\D R^2)= 4(d+2)\D R^{ab}\D R_{ab}-4(\D R)^2-2d \D R^{abcd}\D R_{abcd}\,,
\ee
and at the first order there is no contribution $O(\D R)=0$. Hence, the coefficients that get shifted are
\be
c'_4=c_4-8d_3, \ \ c'_5=c_5+8(d+2)d_3, \ \ c'_6=c_6-4d \ d_3\,,
\ee
while $c_1$ remains unchanged. Putting these values in the expression for $\d T_{\m\n}^{grav}$, we have,
\be
\d T_{\m\n}^{grav}=d \tilde{L}^{d-3}[c'_1+2(d-2)c'_6]h^{(d)}_{\m\n}=d \tilde{L}^{d-3}[c_1+2(d-2)(c_6-4d \ d_3)]h^{(d)}_{\m\n}.
\ee
We can also calculate the holographic stress tensor in \eqref{eq1} directly (see appendix \eqref{holstress}) and show the shift in the coefficient $c_6$ explicitly.

\section{Holographic trace anomalies}
\subsection{For ${\mathcal{L}}(g^{ab},R_{cdef})$}\label{3}
We will now calculate the holographic trace anomalies \cite{weylanom, nojiri} for the Lagrangian in \eqref{lag} following a simple method advocated in Appendix A of \cite{SSS}. This method can be easily implemented on a computer.  Our results will be in agreement with \cite{chinese} wherever we have been able to compare our expressions. We outline the essential steps in the computation of the anomalies.
\begin{enumerate}
\item{ We will first choose a reference background for $g_{(0)ij}$. Since there is no restriction, we can choose any reference background, convenient for the calculation. Note that we can also use multiple reference background for $g_{(0)}$ to determine all the anomaly coefficients.}
\item{The form of $g_{(1)ij}$ is fixed by conformal invariance as \cite{imbimbo}
\be\label{g1}
g_{(1)ij}=-\frac{1}{d-2}(R_{(0)ij}-\frac{R_{(0)}}{2(d-1)}g_{(0)ij})\,,
\ee
where $R_{(0)}$s are constructed out of $g_{(0)}$ respectively.
}
\item{ We will keep $g_{(2)ij}$ arbitrary. Some comments are in order. Demanding the coefficient to $g_{(2)}$ to vanish in $d=4$ in the Lagrangian enforces the condition $c_0=-8 c_1$. This is the same condition as obtained from the lowest order equations of motion. For $d=6$ the relation between $c_0$ and $c_1$ is obtained by demanding that the coefficient of $g_{(3)ij}$ vanishes. We put in $g_{(2)ij}$ for consistency but in the end it does not play a role.}  
\item{ Plugging in the FG expansion in \eqref{FG} into \eqref{lag}, we get
\be
S=\int dz d^d x z^{-d-1}\sqrt{-g_{(0)}}b(x,z),
\ee
where $b(x,z)=b_0(x)+z^2 b_1(x)+\dots$. Next we extract the coefficient of $1/z$ term in the above term which we call $S_{ln}$.}
\item{The trace anomaly in $d$ dimensions is given by
\be
\la T^{\m}_{\m}\ra=b_{d/2},
\ee
where $b_{d/2}$ is the coefficient of $z^d$ in the expansion for $b(x,z)$.}
\item{By matching the term $S_{ln}$ with the expressions for $\la T^{\m}_{\m}\ra$ we can determine various anomaly coefficients.}
\end{enumerate}

\subsubsection{\underline{d=2}}

In $d=2$ the $S_{ln}$ has only one anomaly term which is the Euler anomaly given by $E_2=\frac{1}{4}R$.
Evaluated on the manifold 
\be
ds^2=g_{(0)ij}dx^idx^j= u(\chi^2 dt^2+\frac{d\chi^2}{\chi^2}),
\ee
the Euler anomaly takes the form $E_2=-\frac{1}{2 u}$. The $1/z$ term in the action is given by $S_{ln}=-2c_1$. Equating this with the anomaly term ${\mathcal{A}}=\frac{c}{8\pi}E_2$ and finally putting $u=1$,  we get
\be
c=32\pi\tilde{L} c_1.
\ee

\subsubsection{\underline{d=4}}
In $d=4$ the $S_{ln}$ will contain a linear combination of the Weyl and the Euler anomalies given by
\begin{align}
\begin{split}
E_4&= R_{(0)}^{abcd}R_{(0)abcd}-4R_{(0)}^{ab}R_{(0)ab}+R_{(0)}^2,\\
I_4&=R_{(0)}^{abcd}R_{(0)abcd}-2R_{(0)}^{ab}R_{(0)ab}+\frac{1}{3}R_{(0)}^2,
\end{split}
\end{align}
where $R_{(0)abcd}$ is constructed out of $g_{(0)ab}$. The trace anomaly is given by
\be
\la T^{\m}_{\m}\ra=\frac{c}{16\pi^2} I_4-\frac{a}{16\pi^2} E_4.
\ee
We take $g_{(0)}$ as
\be
g_{(0)ij}dx^i dx^j=u(-\c^2 dt^2+\frac{d\c^2}{\c^2})+v (d\t^2+\sin^2\t \ d\f^2),
\ee
which is of the form $AdS_2\times S_2$. In this background the anomalies take the form 
\be
E_4=-\frac{8}{uv},\ \ \ I_4=\frac{4(u-v)^2}{3 u^2v^2} .
\ee
The coefficient of $1/z$ term in the action is
\be
S_{ln}=\frac{\sin\t}{6uv }(4c_6(u-v)^2+c_1(u^2+4 u v+v^2)).
\ee
Comparing $S_{ln}$ and $\la T^{\m}_{\m}\ra$ we get, after restoring the factors of $\tilde{L}$ in $a$ and $c$
\be
a=2\pi^2 \tilde{L}^3c_1, \ \ c=2\pi^2\tilde{L}^3(c_1+4c_6).
\ee
The $4d$ holographic stress tensor in \eqref{eq1} can thus be written as
\be
\la\d T^{grav}_{\m\n}\ra=4\tilde{L}[c_1+4c_6]h^{(d)}_{\m\n}=\frac{2}{\tilde{L}^2\pi^2} c\ h^{(d)}_{\m\n}.
\ee

\subsubsection{\underline{d=6}}
In $d=6$ there are four anomaly coefficients \cite{deserschwimmer, 6danom} of which three are called the B-type anomalies which are the coefficients of the three Weyl anomalies and the other one is the A-type which is the coefficient of the Euler term in 6d. The trace anomaly in $6d$ is given by
\be\label{tmm}
\la T^{\m}_{\m}\ra=S_{ln}=(\sum_{i=1}^3 B_i I_i+2A E_6)\,,
\ee
where the expressions for the anomalies are given by
\begin{align}
\begin{split}
I_1&= C_{ijkl}C^{imnj}C_{mn}^{ \ \ \ kl},\\
I_2&=C_{ij}^{ \ \ \ kl}C_{kl}^{ \ \ \ mn}C_{mn}^{ \ \ \ ij},\\
I_3&=C_{iklm}(\nabla^2 \d^{i}_{j}+4 R^{i}_{j}-\frac{6}{5}R\d^{i}_{j})C^{jklm},\\
E_0&=384 \pi^3 E_6={\mathcal{K}}_1-12{\mathcal{K}}_2+3{\mathcal{K}}_3+16{\mathcal{K}}_4
-24{\mathcal{K}}_5-24{\mathcal{K}}_6+4{\mathcal{K}}_7+8{\mathcal{K}}_8\,,
\end{split}
\end{align}
where the terms ${\mathcal{K}}_1\dots {\mathcal{K}}_8$ are given by (\ref{Ki}).
To determine the anomaly coefficients we choose for $g_{(0)}$ two manifolds $AdS_2\times S_4 \ \text{and} \ AdS_2\times S_2\times S_2$. In $AdS_2\times S_4$ we have
\be
I_1=-\frac{51(u-v)^3}{100 u^3 v^3},\  I_2=\frac{39(u-v)^3}{25 u^3 v^3}, \ I_3=-\frac{36(19u+v)(u-v)^2}{25 u^3 v^3}, \ E_6=-\frac{144}{u v^2},
\ee
while in the $AdS_2\times S_2\times S_2$ background we have
\begin{align}
\begin{split}
I_1&=-\frac{3(51u^3+21 u^2 v+17 u v^2-17 v^3)}{100u^3 v^3},\ I_2=\frac{3(39u^3-31 u^2 v+13 u v^2+13 v^3)}{25 u^3 v^3},\\ I_3&= -\frac{12(11 u^3-39 u^2 v+17 u v^2+3 v^3)}{25 u^3 v^3},\
E_6=-\frac{48}{ u v^2}.
\end{split}
\end{align}

\par
$S_{ln}$ in the $AdS_2\times S_4$ background takes the form, 
\be\label{sln}
S_{ln}=-\frac{3c_1}{4}-(c_1+8 c_6)\frac{3(u-v)^2}{20 v^2}+(11c_1+94c_6+104\tilde{c}_7-34\tilde{c}_8)\frac{3(u-v)^3}{200 v^3},
\ee
where $\tilde{c}_7 \ \text{and} \ \tilde{c}_8$ are coefficients of the seventh and the eighth term in \eqref{Ki}. Comparing \eqref{sln} and \eqref{tmm}, we get
\be
A=c_1,\ \ B_3=\frac{1}{192}(8c_6+c_1), \ \  68B_1-208 B_2+c_1+20c_6+208\tilde{c}_7-68\tilde{c}_8=0\,.
\ee
Using $AdS_2\times S_2\times S_2$ for $g_{(0)}$ we get one more relation as,
\be 
54 B_1 -24 B_2 +3 c_1 +10 c_6 +24 \tilde{c}_7 -54 \tilde{c}_8=0\,.
\ee
We solve these two equations to get after restoring the factors of $\tilde{L}$,
\be
A=\tilde{L}^5 c_1,\ \ B_3=\frac{\tilde{L}^5}{192}(8c_6+c_1), \ \ 2B_1=\tilde{L}^5(-\frac{c_1}{8}-\frac{c_6}{3}+2\tilde{c}_8), \ \ 2B_2=\tilde{L}^5(-\frac{c_1}{32}+\frac{c_6}{12}+2\tilde{c}_7)\,.
\ee
The holographic stress tensor in \eqref{eq1} can now be re-expressed as,
\be
\la\d T^{grav}_{\m\n}\ra=6\tilde{L}^3[c_1+8c_6]h^{(6)}_{\m\n}=6\tilde{L}^{3}B'_3 h^{(6)}_{\m\n}\,,
\ee
where we define $B'_3=192B_3$. The relation between the holographic stress tensor and the asymptotic metric thus takes the form of \eqref{eq1} where ${\mathcal{C}}_T$ is related to the B-type anomaly coefficient as\footnote{We thank Mark Mezei for point out a mistake in the previous version. The difference arose from missing out a factor of 3 in\eqref{sln}} 
\be
{\mathcal{C}}_T=192f_6 B_3.
\ee

\subsection{ For ${\mathcal{L}}(g^{ab},R_{cdef},\nabla_a R_{bcde})$}\label{4}
We will use the same Lagrangian \eqref{nablar} for the calculation of the holographic anomalies. Here by a scaling argument as in \cite{chinese} it is easy to show that the action with two covariant derivatives acting on two Riemann tensors, will take on the form as in \eqref{nablar}. In the presence of the $\nabla R$ terms in the action, the central charges of the higher derivative theories get modified accordingly. 

\subsubsection{\underline{d=4}}
The additional contribution to the $S_{ln}$ is 
\be
S_{(d)ln}=-\frac{32d_3 \sin\t(u-v)^2}{3uv}\,,
\ee
which combined with the remaining terms give 
\be
S_{ln}==\frac{\sin\t}{6uv}[(4c_6-64d_3)(u-v)^2+c_1(u^2+4 u v+v^2)]\,.
\ee
Comparing these expressions with the usual formula for the anomaly term we get the anomaly coefficients, as 
\be
a=2\pi^2\tilde{L}^3c_1, \ \  \ c=2\pi^2\tilde{L}^3(4c_6+c_1-64d_3).
\ee
We can say that $c'_6=c_6-16d_3$ and hence the holographic stress tensor of \eqref{eq1} becomes 
\be
\la\d T^{grav}_{\m\n}\ra=4\tilde{L}[c_1+4c'_6]h^{(4)}_{\m\n}=\frac{2}{\tilde{L}^2\pi^2}c\  h^{(4)}_{\m\n}\,,
\ee
as before for 4$d$.

\subsubsection{\underline{d=6}}
In 6$d$ the additional contribution to $S_{ln}$ due to the $(\nabla R)^2$ terms in $AdS_2\times S_4$ is,
\be 
S_{(d)ln}=-\frac{36(7u+3v)(u-v)^2}{25 u^2 v }d_3.
\ee
Comparing the total contribution to the coefficient of $1/z$ term with the expression for $\la T^{\m}_{\m}\ra$ for $AdS_2\times S_4 \ \text{and} \ AdS_2\times S_2\times S_2$ we get, after restoring the factors of $\tilde{L}$,
\be
A=\tilde{L}^5c_1,\ \ B_3=\frac{\tilde{L}^5}{192}(c_1+8c_6-192d_3), \ \ 2B_1=\tilde{L}^5(-\frac{c_1}{8}-\frac{c_6}{3}+2\tilde{c}_8+16d_3), \ \ 2B_2=\tilde{L}^5(-\frac{c_1}{32}+\frac{c_6}{12}+2\tilde{c}_7-4d_3)
\ee
where $c'_6=c_6-4d d_3$. The holographic stress tensor in \eqref{eq1} can now be written as
\be
\d T^{grav}_{\m\n}=6\tilde{L}^3[c_1+8c'_6]h^{(6)}_{\m\n}=6\tilde{L}^3 B'_3 h^{(6)}_{\m\n}\,,
\ee
for the 6$d$ case where as before we define $B'_3=192 B_3$. 

\section{Holographic two point function for higher derivative theories in arbitrary dimensions}\label{5}

In this section we will show that the coefficient in the expression for the holographic stress tensor is related to the coefficient in the holographic two point function in arbitrary dimensions for any higher derivative theory whose bulk Lagrangian is of the form ${\mathcal{L}}(g^{ab},R_{bcde},\nabla_a R_{bcde})$. In even dimensions the coefficient of the holographic two point function is related to the coefficient of the two point function in field theory which is proportional to the B-type anomaly coefficient  \cite{osborn}, \cite{erdmenger} (our results in six dimensions are new). The details of the calculation from the field theory side are done in appendix \eqref{A}. In odd dimensions there is no anomaly. We will show that the coefficient appearing in the expression of the holographic stress tensor is related to the coefficient of the holographic two point  functions in arbitrary dimensions.

\noindent As previously, we will consider the action,
\be\label{act}
S=\int d^{d+1}x \sqrt{g}[c_0+c_1\D R+\frac{c_4}{2}\D R^2+\frac{c_5}{2}\D R^{ab}\D R_{ab}+\frac{c_6}{2}\D R^{abcd}\D R_{abcd}]\,,
\ee
where $c_0=-2d c_1$. The advantage of using the above action for  the computation of the two point function is that the result is then valid for any arbirary higher derivative theory of gravity of the form ${\mathcal{L}}(g^{ab},R_{bcde},\nabla_a R_{bcde})$ with $c_6$ replaced by $c_6'$ as argued previously. To compute the two point function it is sufficient to keep upto $O(\D R)^2$ terms only since as we are expanding when we expand around the AdS background, $O(\D R)^3$ terms will start at order $O(h^3)$. 
To compute the two point functions we will follow the arguments of \cite{quasitop} where it is shown that to calculate the two point functions it is sufficient to look at components like $\la T_{xy}T_{xy}\ra$ since the other structures are completely determined by symmetry. Following \cite{quasitop}we turn on a component $r^2h_{xy}(r,z)/L^2$ of the metric perturbations. The quadratic action for the fluctuation of the above form for our case is given by
\beq\label{fluc}
S=\int d^{d+1}x[K_1\phi^2+K_2(\pd_z\phi)^2+K_3\pd_z^2\phi^2+K_4\pd_z^2\phi\pd_r\phi+K_5 (\pd_r\phi)^2+K_6(\pd_r\pd_z\phi)^2\nonumber\\+K_7\pd_r^2\phi\pd_z^2\phi+K_8\pd_r\phi\pd_r^2\phi
+K_9(\pd_r^2\phi)^2+K_{10}\phi\pd_z^2\phi+K_{11}\phi\pd_r\phi+K_{12}\phi\pd_r^2\phi].
\eeq
The last term can be integrated by parts to obtain
\be
K_{12}\phi\pd_r^2\phi=\pd_r(K_{12}\phi\pd_r\phi)-K_{12}(\pd_r\phi)^2-\pd_rK_{12}\phi\pd_r\phi,
\ee
where we have assumed that there exists a generalized Gibbons-Hawking term which takes care of the total derivatives.  We will consider the scalar field to be
\be
\phi(r,z)=\phi_k(r)e^{-i k z}.
\ee
Taking the Fourier transform of the action, after the integration by parts of the last term, we have
\beq
{\mathcal{A}}=\int d^{d+1} k [K_1\phi_k\phi_{-k}+K_2k^2\phi_k\phi_{-k}+K_3 k^4 \phi_k\phi_{-k}-\frac{1}{2}K_4 k^2\phi_k\dot{\phi}_{-k}-\frac{1}{2}K_4k^2\phi_{-k}\dot{\phi}_{k}+K_5\dot{\phi}_k\dot\phi_{-k}\nonumber\\
+K_6k^2\dot{\phi}_k\dot\phi_{-k}-\frac{1}{2}K_7k^2\ddot{\phi}_k\phi_{-k}+\frac{1}{2}\pd_r(K_7k^2\phi_k)\dot{\phi}_{-k}-\frac{1}{2}\pd_r(K_8\dot{\phi}_k)\dot{\phi}_{-k}-\frac{1}{2}\dot{K}_8\dot{\phi}_k\dot{\phi}_{-k}\nonumber\\
-\pd_r(K_9\ddot{\phi}_k)\dot{\phi}_{-k}-K_{10}k^2\phi_k\phi_{-k}+\frac{1}{2}K_{11}\phi_k\dot{\phi}_{-k}+\frac{1}{2}K_{11}\phi_{-k}\dot{\phi}_k-K_{12}\dot{\phi}_k\dot{\phi}_{-k}-\dot{K}_{12}\phi_k\dot{\phi}_{-k}]\,,\nonumber\\
\eeq
where $\dot{}$ denotes derivative with respect to $r$. The terms $K_i$ are given by
\begin{align}
\begin{split}
K_1&=d c_1 r^{d-1}, \ K_2=\frac{3}{2} c_1 r^{d-3}, \ K_3=(\frac{c_5}{4}+c_6)r^{d-5}, \ K_4=\frac{1}{2}[(d+1)c_3+4c_6]r^{d-2},\\
K_5&=\frac{1}{4}[6c_1+(d+1)^2 c_5+4(d+7)c_6]r^{d+1}, \ K_6=2c_6 r^{d-1}, \ K_7=\frac{1}{2} c_5 r^{d-1},\\
K_8&= \frac{1}{2}[(d+1)c_5+12c_6]r^{d+2}, \ K_9=\frac{1}{4}(c_5+4c_6)r^{d+3}, \ K_{10}=2c_1r^{d-3}, \ K_{11}=2(d+2)c_1 r^d, \\ 
K_{12}&=2c_1 r^{d+1}\,.
\end{split}
\end{align}
After integration by parts  the above action can be written as a boundary term \footnote{We have assumed that the volume counterterm gets rid of $\phi_k \phi_{-k}$ terms as in \cite{liutseytlin, holGB}.}
\begin{align}
\begin{split}
\pd{\mathcal{A}}&=-\frac{1}{2}K_4k^2\phi_k\phi_{-k}+K_5\dot{\phi}_k\phi_{-k}+K_6k^2\dot{\phi}_k\phi_{-k}+\frac{1}{2}\pd_r(K_7k^2\phi_k)\phi_{-k}-\frac{1}{2}\dot{K}_8\dot{\phi}_k\phi_{-k}\\
&-\pd_r(K_9\ddot{\phi}_k)\phi_{-k}-K_{12}\dot{\phi}_k\phi_{-k}\,,
\end{split}
\end{align}
where again $\dot{}$ denotes derivative with respect to $r$. The solution to $(\Box+2)h_{ab}=0$ still solves the higher derivative equations\footnote{See e.g.\cite{deser}, alternatively we just assume that there is a massless graviton which by definition solves this equation.}.
The solution  is given by (restoring the AdS radius $\tilde L$)
\be
\phi_k(r)=\frac{2\tilde{L}^4|k|^{d/2}}{d r^{d/2}}K_{d/2}(\frac{\tilde{L}^2 |k|}{ r}),
\ee
where $K_{d/2}$ is the modified Bessel function of the second kind. The normalization constant is obtained by imposing the condition that $\phi_k(r=\infty)=1$ and $d$ is the field theory dimension. By plugging this solution back into the surface term $\pd{\mathcal{A}}$ and extracting the coefficient of $k^d$ term in the resulting expression, we get, for $AdS_{d+1}/CFT_d$ after restoring the factors of $\ell_p$,
\be
\la T_{ab}(x)T_{cd}(x')\ra=\frac{1}{\tilde{L}^2}{\mathcal{C}}_T \frac{{\mathcal{I}}_{ab,cd}(x-x')}{|x-x'|^{2d}}\,,
\ee
where the coefficient ${\mathcal{C}}_T$ is given by
\be
{\mathcal{C}}_T=f_d\tilde{L}^{d-1}[c_1+2(d-2)c_6],
\ee
where $f_d$ is the constant factor given by
\be
f_d=2\frac{d+1}{d-1}\frac{\Gamma[d+1]}{\pi^{d/2}\Gamma[d/2]}.
\ee
Thus the expression for $\la T_{\m\n}^{grav}\ra$ in \eqref{eq1} becomes,
\be
\la T_{\m\n}^{grav}\ra=\frac{d}{f_d\tilde{L}^2}{\mathcal{C}}_T h^{(d)}_{\m\n}\,.
\ee
Note that while we have assumed the existence of a suitable generalized Gibbons-Hawking term we have not used counterterms involving boundary curvature tensors in our calculations. We have explicitly checked, the addition of such counterterms will not alter our findings. 
\section{Three point functions}\label{6}
The fact that we were able to get the one point and two point functions from the background field expansion seems to suggest that the analysis can be extended to the calculation of 3-point functions using the same technique.  We will carry out the analysis first by considering a higher derivative Lagrangian of the form ${\mathcal{L}}(g^{ab},\D R_{abcd})$ and then extending the analysis to the case where ${\mathcal{L}}(g^{ab},\D R_{abcd},\nabla_a \D R_{bcde})$. 

Direct holographic calculation of the three point functions are involved and challenging. We will follow the alternative route to derivation of the three point functions following the analysis of \cite{HM} and used in \cite{HM, quasitop, edelstein}. The energy flux associated with a localized perturbation of fixed energy $\e_{ij}T^{ij}$, where $\e_{ij}$ is the polarization tensor, in the $d(>3)$ dimensional CFT background is given by
\be
\la \epsilon(\textbf{n})\ra=\frac{E}{4\pi\Omega_{d-2}}[1+t_2(\frac{\e_{ij}^* \e_{ik}n^jn^k}{\e_{ij}^* e_{ij}}-\frac{1}{d-1})+t_4(\frac{|\e_{ij}n^jn^k|^2}{\e_{ij}^* e_{ij}}-\frac{2}{d^2-1})].
\ee
Here $E$ is the total energy flux, $\textbf{n}$ is the outward normal in the direction in which the flux is measured and $\Omega_{d-2}$ is the volume of a unit $S^{d-1}$ sphere. The coefficients $t_2\ \text{and} \ t_4$ are determined holographically in the following way. From the holographic side we insert graviton perturbations $h_{\m\n}$ dual to the energy insertion in the field theory and evaluate the on-shell cubic term in the higher derivative Lagrangian corresponding to these graviton insertions. Following \cite{HM, quasitop}, we consider the shockwave background with perturbations in $d$ dimensions:
\be
ds_{sw}^2=\frac{\tilde{L}^2}{u^2}[\d(y^+)W(\vec{y},u)(dy^+)^2-dy^+dy^-+d\vec{y}^2+du^2]+h_{ij}dx^idx^j\,,
\ee
where $\vec{y}^2=\sum_{i=0}^{d-2}y_i^2$ and $d$ is the dimension of the field theory. The function $W(\vec{y},u)$ is given by
\be
W(\vec{y},u)=\frac{2^{d-1}}{(1+n_{d-1})^{d-1}}\frac{u^d}{(u^2+(\vec{y}-\vec{Y})^2)^{d-1}},
\ee
where $n_{d-1}$ is the $(d-1)$th component of the normal vector given by 
\be
n_{d-1}= (1-n_i^2)^{\frac{1}{2}},\ \text{and} \  Y^i=\frac{n^i}{1+n_{d-1}}.
\ee
$W$ satisfies the following equation in any higher derivative theory of gravity\cite{horowitz}, 
\be
\pd_u^2W-\frac{d-1}{u}\pd_uW+\sum_{i=1}^{d-2}\pd_{y^i}^2W=0.
\ee
The transverse traceless gauge brings down the number of independent components of the perturbations. In $d$ dimensions we can consider the perturbation of the form $h_{y^1 y^2}=\tilde{L}^2/u^2\phi(\vec{y},u),$
while $h=0= \nabla^\m h_{\m\n}$ relates the other components as 
\be
\pd_-h_{y^+y^1}=\frac{1}{2}\pd_{y^2}h_{y^1y^2}\,, \ \pd_-h_{y^+y^2}=\frac{1}{2}\pd_{y^1}h_{y^1y^2}\,,\
\pd_-h_{y^+y^+}=\frac{1}{4}(\pd_{y^1}h_{y^+y^1}+\pd_{y^2}h_{y^+y^2}).
\ee
It is sufficient to turn on these components only for general $d(>3)$ dimensions. The component $h_{y^1y^2}$ satisfies the lowest order equation of motion for a scalar field in the $AdS_{d+1}$ background given by,
\be
\pd_u^2\phi-\frac{d-1}{u}\pd_u\phi+\sum_{i=0}^{d-2}\pd_{y^i}^2\phi-4\pd_+\pd_-\phi=0.
\ee
\subsection{${\mathcal{L}}(g^{ab},  R_{cdef})$}
Using the equation of motion for $\phi \ \text{and} \ W$ we can evaluate the on-shell cubic effective action to get the most general form in $d(>3)$ dimensions as\footnote{To reach this simple form, we need to integrate by parts and use the on-shell conditions multiple number of times.}
\begin{align}\label{3pt}
\begin{split}
S^{(3)}_{W\phi^2}&=-\frac{1}{4}\int d^{d+1}x\sqrt{-g}\phi\pd_-^2\phi \bigg[2(c_1+2(d-2)c_6)W-2u(2c_6-12d\tilde{c}_7+3(3d-4)\tilde{c}_8)\pd_uW\\
&-24u^2(2\tilde{c}_7-\tilde{c}_8)\sum_{i>2}^{d-1}\pd_{i-2}^2W+u^2(2c_6-12(8-d)\tilde{c}_7+3(12-d)\tilde{c}_8)(\pd_1^2W+\pd_2^2W)\\
&-24u^3(2\tilde{c}_7-\tilde{c}_8)(\sum_{i=1}^{d-1}\pd_i^2\pd_uW-u\sum_{i>j}^{d-1}\pd_i^2\pd_j^2W)\bigg]{\bigg |}_{u=1,y_1=0,y_2=0}\,.
\end{split}
\end{align}
Note that the integral localizes on $u=1,y_1=0,y_2=0$ \cite{HM,hofman,quasitop}. As a result we do not have to worry about boundary terms like the generalized Gibbons-Hawking term or the boundary counterterms in this calculation.
Comparing with the standard form given in \cite{quasitop},
\be
S^{(3)}_{W\phi^2}=-\frac{{\mathcal{C}}_T}{4f_d\tilde{L}^{d-1}}\int d^{d+1}x \sqrt{-g} \ \phi\pd_-^2\phi \ W[1+t_2 T_2+t_4 T_4],
\ee
and $T_2$ and $T_4$ are given by
\be
T_2=\frac{n_1^2+n_2^2}{2}-\frac{1}{d-1}\,,\ T_4=2n_1^2n_2^2-\frac{2}{d^2-1}\,,
\ee
while the coefficients $t_2,t_4$ are given by \footnote{If we set $W=1$ then we would be left with just the two point function which would be proportional to $\mathcal{C}_T$ as expected.},
\be\label{t2t4}
t_2=\frac{d(d - 1)}{c_1+2(d-2)c_6}[2c_6-12 (3d + 4) \tilde{c}_7+3 (7 d + 4) \tilde{c}_8], \ 
t_4= \frac{6d(d^2 - 1)(d+2)}{c_1+2(d-2)c_6}(2\tilde{c}_7 -\tilde{c}_8).
\ee
This is the expected result for cubic Lovelock theory \cite{edelstein} where $2\tilde c_7=\tilde c_8$ and hence $t_4=0$. We have also checked that our general expressions are in agreement with \cite{quasitop, bpk}.

\subsection{${\mathcal{L}}(g^{ab}, R_{cdef},\nabla_a R_{bcde})$}
We now extend the analysis of the previous section to higher curvature theories containing covariant derivatives of the Riemann tensor. In section \eqref{nabla} we have shown how the presence of the $\nabla^2\D R^2$ terms modify the coefficient $c_6\rightarrow c'_6=c_6-4d d_3$. In addition the cubic order coefficients are modified as
\be
\tilde{c}_7\rightarrow \tilde{c}'_7=\tilde{c}_7-d_3 \ \tilde{c}_8\rightarrow \tilde{c}'_8=\tilde{c}_8+4d_3, \ \tilde{c}_2\rightarrow \tilde{c}'_2=\tilde{c}_2-4d_3, \ \tilde{c}_5\rightarrow \tilde{c}'_5=\tilde{c}_5+4d_3, \ \tilde{c}_6\rightarrow \tilde{c}'_6=\tilde{c}_6+2d_3.
\ee
Thus $c_i \ \text{and} \ \tilde{c}_i$ in \eqref{t2t4} will be replaced by $c'_i$  and $\tilde{c}'_i$ respectively. In this section we will consider additional terms like $\nabla^2 \D R^3$ terms in the action \eqref{lag}.  For $\nabla\nabla \D R^3$ terms, since the linearized Ricci tensor and scalar curvature vanishes by using the tracelessness condition and the lowest order equation of motion satisfied by $h_{ab}$, as shown in section \eqref{nabla}, the terms which contribute to the three point functions are 
\be
S_3=e_1 R^{ab \ \ }_{ \ \ cd} R^{cd \ \ }_{ \ \ ef}\nabla^2 R^{ef \ \ }_{ \ \ ab}+e_2  R^{a \ b \ }_{ \ c \ d} R^{c \ d \ }_{ \ e \ f}\nabla^2 R^{e \ f \ }_{ \ a \ b}.
\ee
To show that these are the only tensor structures that contribute to the three point functions, consider the first term which can be shown to be, 
\be
R^{ab \ \ }_{ \ \ cd} R^{cd \ \ }_{ \ \ ef}\nabla^2 R^{ef \ \ }_{ \ \ ab}=\nabla_m(\D R^{ab \ \ }_{ \ \ cd}\D R^{cd\ \ }_{ \ \ ef}\nabla^m\D R^{ef \ \ }_{ \ \ ab})-2\nabla_m\D R^{ab \ \ }_{\ \ cd}\nabla^m\D R^{ef \ \ }_{\ \ ab}\D R^{cd \ \ }_{ \ \ ef},
\ee
where the overall factor of 2 comes because of $\nabla$ acting on any term other than $\nabla \D R$ are equivalent.  Similarly it can be shown for the second term as well. 

These terms have additional contribution to the coefficients $t_2 \ \text{and} \ t_4$ but ${\mathcal{C}}_T$ remains unaffected. The coefficients $c_i$ and $\tilde{c}_i$ in \eqref{t2t4} are replaced by their effective values as, 
\be
t_2=\frac{d(d - 1)}{c_1+2(d-2)c'_6}[2c'_6-12 (3d + 4) \tilde{c}''_7+3 (7 d + 4) \tilde{c}''_8], \ 
t_4= \frac{6d(d^2 - 1)(d+2)}{c_1+2(d-2)c'_6}(2\tilde{c}''_7 -\tilde{c}''_8)\,,
\ee
where $\tilde{c}''_7=\tilde{c}'_7+2d e_1 \ \text{and} \ \tilde{c}''_8=\tilde{c}'_8+2d e_2$. 

We mention here that although we leave the analysis for the general $\nabla\dots\nabla\D R\dots\D R$ terms for future work, we feel that this pattern will continue to persist so that the $\nabla$ terms in the action \eqref{lag} will modify the coefficients appearing in the two and the three point functions and the form of $C_T, \ t_2, \ \text{and} \ t_4$ will remain the same as in \eqref{t2t4} with the coefficients being replaced by similar shifted ones as discussed above.

\section{ Application: $\eta/s$ for higher derivative theories}\label{ebs}
As an application of the background field expansion method, we calculate the ratio of the shear viscosity and entropy density \cite{KSS} for higher derivative theories \cite{etabs}. This can be done in arbitrary dimensions but for simplicity, we will illustrate for the $d=4$ plasma. Following \cite{miguel}\footnote{See also \cite{db}.} we will use the pole method to calculate the shear viscosity where only the near horizon data is important. Following \cite{miguel} we write the black hole metric as
\be
ds^2=\frac{\tilde{L}^2}{4f(z)}\frac{dz^2}{(1-z)^2}+\frac{r_0^2}{\tilde{L}^2(1-z)}[-f(z)dt^2+(dx_1+\phi(t) dx_2)^2+dx_2^2+dx_3^2].
\ee
To compute the shear viscosity and the entropy density we need to construct the horizon perturbatively by solving the equations of motion for the higher derivative action order by order in coordinate distance from the horizon but exactly in the couplings. The solution can be written as 
\be
f(z)=2z+f_2 z^2+f_3 z^3+\dots,
\ee
where $f_2$ and $f_3$ are functions of the coefficients appearing in the action. The factor of $2$ fixes the temperature with a particular normalization as
\be
T=\frac{r_0}{\pi \tilde{L}^2}\,.
\ee
To compute the shear viscosity we have to plug in a perturbation $h_{xy}$ and compute the retarded Green's function 
\be
G_R^{xy,xy}(\omega)=-i\int \theta(t)\la T^{xy}(t)T^{xy}(0)\ra e^{-i\omega t}\,,
\ee
and finally
\be
\eta=\lim_{\omega\rightarrow0}\frac{\text{Im} G_R^{xy,xy}(\omega)}{\omega}.
\ee
We plug in the perturbation corresponding to the shear mode at zero momentum corresponding to the change of basis 
\be
dx_1\rightarrow dx_1+\phi(t)dx_2.
\ee
Plugging this into the action \eqref{lag}, we get
\be
S_{\phi^2}=\int d^5x ({\mathcal{A}}_1 \phi'_{\omega}\phi'_{-\omega}+{\mathcal{A}}_2 \phi''_{\omega}\phi''_{-\omega}),
\ee
where ${\mathcal{A}}_1 \ \text{and} \ {\mathcal{A}}_2$ are function of the coefficients in the action \eqref{lag}. Following \cite{quasitop, miguel}, we apply the pole method for any general action of the form
\be
S_{\phi^2}=\int d^dx dz {\mathcal{L}}^{(2)}_{\phi}(\pd_z\phi,\pd_t\phi),
\ee
using which
\begin{align}
\begin{split}
\eta&=-8\pi T\lim_{\omega\rightarrow0}\frac{Res_{z=0}{\mathcal{L}}^{(2)}_{\phi=z^{i\omega T}}}{\omega^2} .
\end{split}
\end{align}

Putting in $\phi(t)=e^{-i\omega t}$ we thus extract the coefficient of $1/z$ term and expanding upto quadratic orders in $\omega$, we finally get, 
\be
\eta=r_0^3(A_1+B_1 f_2+C_1 f_2^2+C_3 f_3),
\ee
where the coefficients $A_1, B_1, C_1$ and $C_3$ are functions of the coefficients in  \eqref{lag}. Similarly the entropy density for the higher derivative action is computed using the Wald formula and takes on the form
\be
s=4\pi r_0^3(A_2+ B_2 f_2+C_2 f_2^2).
\ee
Note that in the above expressions for both $\eta$ and $s$, we have set the AdS radius $\tilde{L}=1$. The deviation of the $\eta/s$ ratio from the KSS bound \cite{KSS} for the action \eqref{lag} corresponding to the case when $t_4=0$ and in the absence of $O((\Delta R)^3)$ terms is simply given by
\be
(\frac{\eta}{s}-\frac{1}{4\pi})s=-2c_6r_0^3.
\ee 

The explicit form of $\eta$ and $s$ are given in the appendix \eqref{etas} for a general $R^2$ theory where it is shown that for particular values of the coupling constants of the general $R^2$ theory, the ratio can be driven to zero. As another example we quote the results for the $W^3$ gravity below where the lower bound for $\eta/s$ is much lower than the KSS bound. \\

\textbf{Example: $W^3$ theory}

\begin{figure}[ht]\label{fig}
\begin{center}
\includegraphics{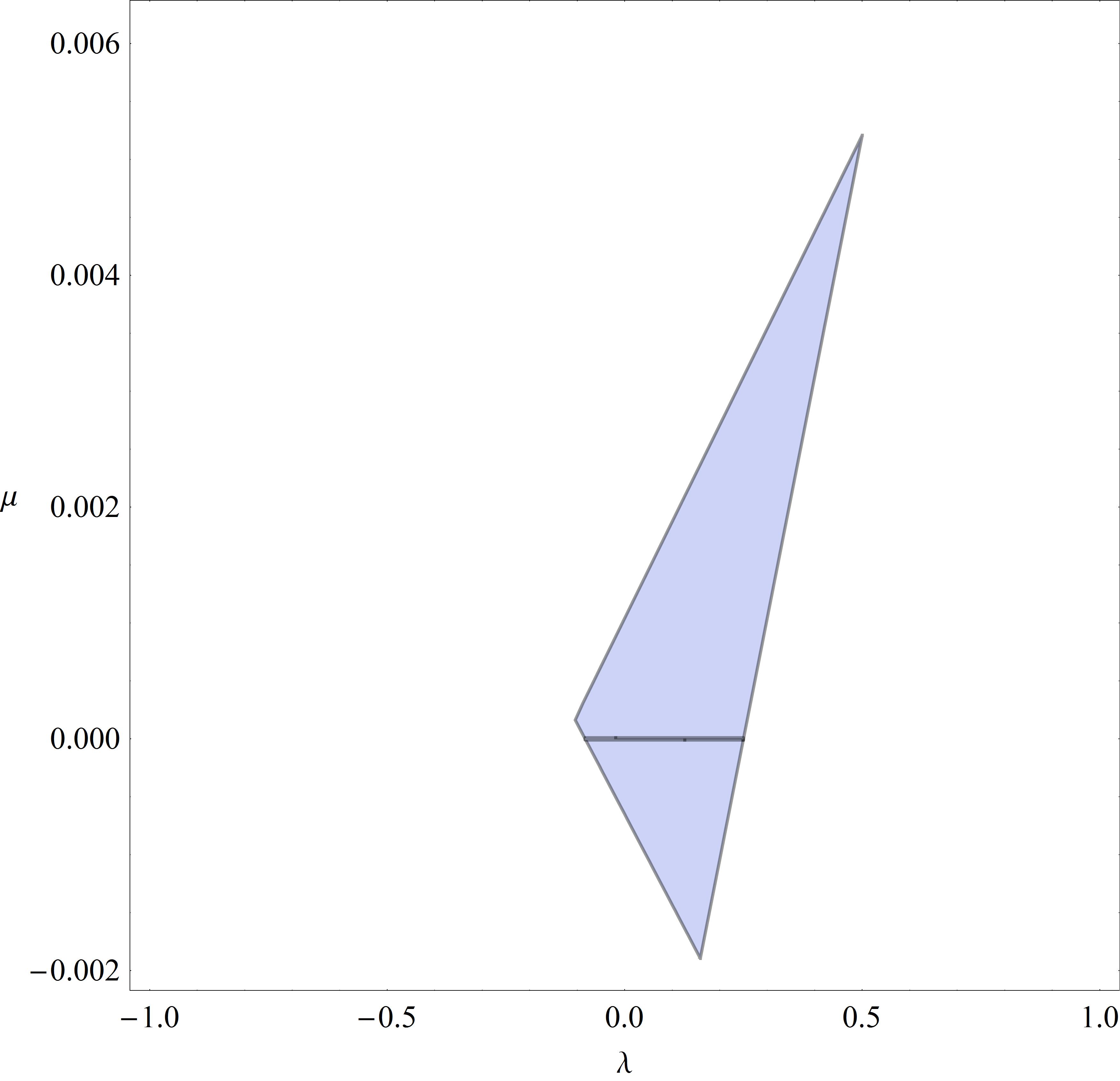}
\caption{$\l$ vs $\m$ plot. The horizontal line corresponds to $\m=0$. $(\eta/s)_{min}=0.55/4\pi$ for $\m=0$ and $(\eta/s)_{min}=0.17/4\pi$ for $\m\neq 0$.}
\end{center}
\end{figure}

In \cite{quasitop}, a specific six derivative theory was considered which led to equations of motion for fluctuations which were second order in radial derivatives. The motivation was to consider putting bounds on $\eta/s$ using the positive energy constraints as well as comparing these with the causality constraints. It was found that the positive energy constraints bounded the couplings and led to $(\eta/s)_{min}\approx 0.414/4\pi$. In light of our general analysis, we will consider the following six derivative Lagrangian \cite{HM} which also leads to $t_4\neq 0$ and we will put bounds on the couplings. \cite{HM} had already considered this action perturbatively in the couplings:
\be
S=\int d^5x \sqrt{g} [R+\frac{12}{\tilde{L}^2}+\frac{\tilde{L}^2}{2}\l W^2+\tilde{L}^4 \m W^3]\,,
\ee
where $W^2=C_{abcd}C^{abcd}$ and $W^3=C^{ab}_{cd}C^{cd}_{ef}C^{ef}_{ab}$ with $C_{abcd}$ being the Weyl tensor. If we expand this action around the $AdS$ background to get \eqref{lag}, then the coefficients of \eqref{lag} for this action are given by
$
c_0=-8c_1, c_1=1, \ c_4=\frac{\l}{6}, \ c_5=-\frac{4\l}{3},\ c_6=\l, \tilde{c}_1=\frac{\m}{2}, \tilde{c}_2=-4\m,\tilde{c}_3=\frac{8\m}{3}, \tilde{c}_4=\frac{64\m}{27},\
 \tilde{c}_5=-\frac{14\m}{9}, \ \tilde{c}_6=\frac{7\m}{54}, \tilde{c}_7=\m \ \text{and} \  \tilde{c}_8=0
$.
Note that for $W^3$ gravity $\fin=1$ and $\tilde{L}=1$. The coefficients ${\mathcal{C}}_T, t_2$ and $t_4$ take the form
\be
{\mathcal{C}}_T=2(1+4\l), \  t_2=\frac{24(\l-96\m)}{1+4\l}, \ t_4=\frac{4320\m}{1+4\l}.
\ee
Using the constraints for $t_2\ \text{and} \ t_4$ listed in \cite{quasitop}, we find that $\l \ \text{and} \ \m$ are bounded (see figure\eqref{fig}). The shear viscosity and the entropy density for this action takes the form
\begin{align}
\begin{split}
\eta&= \frac{r_0^3}{6}[3-6(1+2f_2)\l-16(7-40f_2+16f_2^2+36f_3)\m],\\
s&=\frac{2\pi r_0^3}{3}[3+6(1-2f_2)\l+16(1-2f_2)^2\m]\,,
\end{split}
\end{align}
where $f_3$ is given in terms of $f_2$ by,
\be
f_3=\frac{270-64\m+18\l+f_2(216-171\l+656\m)-6f_2^2(9-42\l+304\m)+4f_2^3(9\l+368\m)+128f_2^4\m}{36(-9-6(1-2f_2)\l+16(1-2f_2)^2\m)},
\ee
where $f_2$ satisfies 
\be
64(1-2f_2)^3\m+36(6(1+f_2)-\l+4(1-f_2)f_2\l)=0.
\ee
This equation has three roots and we will choose the correct root as the one which for the Einstein case goes to $f_2=-1$. Substituting for the Einstein value of $f_2$ we also get that $f_3=0$ in the Einstein limit. We present the bounds on $\l\ \text{and}\ \m$ in Fig.\eqref{fig} obtained from the causality constraints given by \cite{quasitop},
\begin{align}
\begin{split}
\text{Tensor:}&\ 1-\frac{1}{3}t_2-\frac{2}{15}t_4\geq0,\\
\text{Vector:}&\ 1+\frac{1}{6}t_2-\frac{2}{15}t_4\geq0,\\
\text{Scalar:}&\ 1+\frac{1}{3}t_2+\frac{8}{15}t_4\geq0.
\end{split}
\end{align}

The minimum values of $\eta/s$ for $\m\neq0$ lie close to the uppermost vertex of the triangle. For Weyl squared gravity $\m=0$ and constraints give $-1/12<\l<1/4$. This is  presented as a single line interval in the $\l-\m$ plot. The minimum value of $\eta/s$ corresponds to $\l=1/4 \  \text{and}\ \m=0$ which is at the extreme right end of the interval.  
The minimum value of $\eta/s$ for the $W^3$ gravity is given by
\be
\frac{\eta}{s}\approx\frac{0.17}{4\pi},
\ee
for $\l=1/2, \ \text{and}\ \m=1/192$ which is the uppermost vertex of the triangle. For $\m=0$, \textsl{i.e.,} for Weyl squared gravity the minimum value of the ratio is $\eta/s\approx0.55/4\pi$.

Thus even though $\la \epsilon\ra>0$ for general $R^2$ theory, the $\eta/s$ ratio can be driven to zero as we show in appendix \eqref{etas}. Further for $W^2$ theory we can see that the bound goes down to about 55\% of the KSS value whereas for $W^2+W^3$ theory it is 17\% of the KSS bound. 
There are non-unitary modes in this theory. So it appears that unitarity is not a prerequisite for a bound. As in \cite{quasitop}, there could be potential plasma instabilities and it may be interesting to analyse these.

\section{Discussion}\label{7}
In this paper we have computed one, two and three point functions for a general gravity Lagrangian of the form ${\mathcal{L}}(g^{ab}, R_{cdef},\nabla_a R_{bcde})$. We explained that the coefficient appearing as the proportionality between the renormalized stress tensor and the bulk metric is related to ${\cal C}_T$, the coefficient appearing in the two point function of stress tensors. Further we saw how this relates to B-type anomaly coefficients in even dimensions.  We also computed three point functions for bulk Lagrangians of the above form in arbitrary dimensions.

Our general form of the action given in eq. (\ref{lag}) packages the A-type anomaly coefficient (or its analog in odd dimensions) into $c_1$ while ${\cal C}_T$ is given in terms of $c_1, c_6$. Again we emphasise that all these coefficients themselves depend on {\it all} higher derivative terms that appear in the original bulk Lagrangian. This simple separation of the A-type anomaly coefficient as a proportionality constant in front of $\Delta R$ makes it very tempting to think that this is a useful starting point for a general proof of holographic version of the $a$-theorem \cite{athm} in arbitrary dimensions. We can speculate how this may work: First note that the background around which we are expanding could be either the AdS in the ultraviolet or the AdS in the infrared. This means that the respective background expanded Lagrangians must be equal to one another. If there was a matter sector as well, it makes sense to do a background expansion of this sector where we will use the background for the matter fields to be their values in AdS, for example for  a scalar field this will be a constant (different constants in the UV and IR). We thus have a natural separation between the gravity sector and the matter sector--this was one of the vexing issues in the current literature on holographic c-theorems \cite{cthm}; namely how does one define any energy condition if matter couples to the higher curvature terms. Thus we can envisage a situation where on the LHS we have a term proportional to $(a_{UV}-a_{IR})R$ plus other curvature terms while on the RHS we can place the difference between the UV and IR matter Lagrangians. It is very tempting to speculate that $(a_{UV}-a_{IR})>0$ is necessary for there to be no non-unitary modes on the LHS arising from expanding $R$ which in turn is necessary (but may not be sufficient) so that there are no non-unitary modes in the matter Lagrangian. It will be nice to work this out in complete detail as this will shed light on how the proof of the $a$-theorem may work in arbitrary dimensions.

Another important question is to extend our methods and results to four point stress tensor correlation functions. As we pointed out in the introduction, while the A-type trace anomaly in 4d is related to two point and three point functions, in higher dimensions it appears to depend on higher point correlation functions. Also in 3d since there is no analog of $t_2$, it is unclear if the analog of the A-type trace anomaly (proportional to $c_1$) can be extracted from local correlation functions at all--this appears to be consistent with recent claims in \cite{jaff}. The general forms for $t_2$ and $t_4$ that we have derived also seem to suggest that in order to relate the A-type anomaly coefficient in dimensions higher than 4 to the coefficients appearing in correlation functions will need at least four point functions. Furthermore, it could well be that the coupling constants for higher derivative theories are further constrained by considering four point functions\footnote{It will also be interesting to compare how constraints from entanglement entropy \cite{rel} compare with these ones.}. These reasons are sufficient motivation to look at the four point functions in the general gravity theories we have considered in this paper. May be the techniques developed in \cite{raju} could help us out here.

It will be interesting to extend our results to completely general bulk Lagrangians of the form ${\mathcal L}(g^{ab}, R_{cdef},\nabla_a R_{bcde}, \nabla_{(a}\nabla_{b)}R_{cdef},\cdots)$. We expect that for the one, two and three point functions, the simple features we have found in this paper will continue to hold. Finally, it should be possible to extend our methods to study correlation functions which involve the massive graviton modes and $T_{\mu\nu}$ \cite{grumiller}.

\section*{Acknowledgments}
We thank Shamik Banerjee, Johanna Erdmenger, Daniel Grumiller, Rob Myers, Miguel Paulos and the theory groups at TIFR and HRI for useful discussions. 
KS thanks Daniel Grumiller and his group in Vienna for hospitality during the course of this work. AS thanks TIFR, HRI and YITP (especially Tadashi Takayanagi) for hospitality during the course of this work.

\appendix

\section{ The Lagrangian in terms of \cite{FM}}\label{A1}
Consider the background field expanded Lagrangian given by
\beq
{\mathcal{L}}=a_0+a_2\D R+ b_1 \D R^2+b_2 \D R^{ab}\D R_{ab}+b_3\D R^{abcd}\D R_{abcd}+\sum_{i=1}^{8}\tilde{c}_i\D {\mathcal{K}}_i\nonumber\\+Z^{efabcdmprs}\nabla_e\nabla_f\D R_{mprs}\D R_{abcd}+\dots.
\eeq
The Wald functional for any gravity dual following \cite{IW} is,
\be
E_{R}^{abcd}=\frac{\pd {\mathcal{L}}}{\pd R_{abcd}}-\nabla_{e}(\frac{\pd {\mathcal{L}}}{\pd \nabla_{e}R_{abcd}})+\cdots\,,
\ee
which for the above Lagrangian takes the form
\be 
E_{R}^{abcd}=a_2 g^{\la ab}g^{cd\ra}+Y^{abcdefgh}\D R_{efgh}-Z^{efabcdmprs}\nabla_{e}\nabla_{f}\D R_{mprs}+\cdots,
\ee
where $Y^{abcdefgh}$ is the tensor structure that comes from the second order terms in $\D R$ and $Z^{efabcdmprs}$ comes from the $\nabla_aR_{bcde}$ terms in the Lagrangian.
To connect this to eq.(6.8, 6.9, 6.11) of \cite{FM} we need to evaluate $E_R^{abcd} \ \text{and} \ \d E_R^{abcd}$ around AdS space. We split the background metric as $g_{ab}=g_{(0)ab}+\D g_{ab}$. Then $\bar{R}_{abcd}$ can be written as
\be
\bar{R}_{abcd}=-(g_{ac}g_{bd}-g_{ad}g_{bc})=-(g_{(0)ac}g_{(0)bd}-g_{(0)ad}g_{(0)bc})-(g_{\la(0)ab}\D g_{cd\ra})=R^{(0)}_{abcd}-(g_{\la(0)ab}\D g_{cd\ra})\,,
\ee
and
\be
\D R_{abcd}=R_{abcd}-\bar{R}_{abcd}=\D^{\!0} R_{abcd}+g_{\la(0)ab}\D g_{cd\ra}\,,
\ee
where $\D^{\!0}R_{abcd}=R_{abcd}-R^{(0)}_{abcd}$ is the expansion around the background with only $g_{(0)ab}$. Using this relation with the fact that 
\begin{align}
\begin{split}
Y^{abcdefgh}&=Y_{0}^{abcdefgh}+O(\D g),\\
Z^{efadcdmprs}&=Z_{0}^{efabcdmprs}+O(\D g),
\end{split}
\end{align}
where $Y_0$ and $Z_0$ denote the quantities calculated with the metric $g_{(0)}$ which is the AdS metric for our purpose. To begin with, we consider an action without the $\nabla R$ terms. Then
\be
E_{R}^{abcd}=a_2 g^{\la ab}g^{cd\ra}+Y_0^{abcdefgh}\D^0 R_{efgh}+O(\D g),
\ee
where we have used the tensor structure of $Y$ as
\be
Y^{abcdefgh}=b_1 g^{ac}g^{bd}g^{eg}g^{fh}+b_2 g^{ac}g^{eg}g^{bf}g^{dh}+b_3 g^{ae}g^{bf}g^{cg}g^{dh},
\ee
and when evaluated around the AdS background we have at the leading order
\be
E_{R}^{abcd}=a_2 g_{0}^{\la ab}g_{0}^{cd\ra}.
\ee
Comparing with eq(6.8) of \cite{FM} we have $a_2=c_1$.
Next we compute
\be
\frac{\pd E^{abcd}_{R}}{\pd g^{ef}}=2c_1 h^{\la ab}g^{cd\ra}+\frac{\pd}{\pd g^{ef}}(Y \D R).
\ee

The last term gives around the AdS background 
\be
\frac{\pd}{\pd g^{ef}}(Y \D R)=\frac{\pd Y}{\pd g}\D^{\!0} R+\frac{\pd}{\pd g^{mn}}(Y_0^{abcdefgh}g_{\la(0)ef}\D g_{gh\ra}).
\ee  
The first term vanishes when evaluated on AdS and thus
\be 
\frac{\pd }{\pd g^{ef}}(Y\D R)|_{AdS} \ h^{ef}=2(2d b_1+b_2) h g^{\la ab}g^{cd\ra}+2((d-1)b_2+4b_3) h^{\la ab}g^{cd\ra}.
\ee 

Further 
\be
\frac{\pd E_{R}^{abcd}}{\pd R^{efgh}}|_{AdS} \ \d R_{efgh}=Y_0^{abcdefgh}\d R_{efgh}\nonumber
=b_1{\mathcal{R}}g^{\la ab}g^{cd\ra}+b_2{\mathcal{R}}^{\la ab}g^{cd\ra}+b_3{\mathcal{R}}^{abcd},
\ee
Comparing with eq (6.11) of \cite{FM} we get $b_1=c_4/2, \ b_2=c_5/2, \ b_3=c_6/2$. Further,
\be
c_2=-2d c_4-c_5,\ \ \ c_3=2c_1-(d-1)c_5-4c_6.
\ee
The Lagrangian \eqref{lag} thus can be written as, 
\be\label{action1}
S=\int d^{d+1}x \sqrt{g}[c_0+c_1\D R+\frac{c_4}{2} \D R^2+\frac{c_5}{2} \D R^{ab} \D R_{ab}+\frac{c_6}{2} \D R^{abcd}\D R_{abcd}+\sum_{i=1}^{8}\tilde{c}_i\D {\mathcal{K}}_i+Z\D R\nabla\nabla\D R+\cdots]\,.
\ee

\section{ Details of calculation for section \eqref{nabla}}\label{bianchi}

The Bianchi identity reads 
\be
\nabla_aR_{bcde}+\nabla_bR_{cade}+\nabla_cR_{abde}=0\,.
\ee
Then
\be
\nabla^2 R_{bcde}=\nabla^a\nabla_bR_{acde}-\nabla^a\nabla_cR_{abde}.
\ee
Using 
\be
\nabla^a\nabla_bR_{acde}=\nabla_b\nabla^aR_{acde}+R^f_bR_{acde}+R^{a \ \ f}_{ \ bc \ }R_{afde}+R^{a\ \ f}_{ \ bd \ }R_{acfe}+R^{a \ \ f}_{ \ be \ }R_{acdf},
\ee
we have
\be
R^{bcde}\nabla^2R_{bcde}=2R^{bcde}\nabla_b\nabla^aR_{acde}+2R^{bcde}R^f_bR_{fcde}
+2R^{bcde}R^{a \ \ f}_{ \ bc \ }R_{afde}+4R^{bcde}R^{a\ \ f}_{ \ bd \ }R_{acfe}.
\ee
Again using the Bianchi Identity,
\be
\nabla^aR_{acde}=\nabla_dR_{ce}-\nabla_e R_{cd}\,,
\ee
we can write, neglecting the total derivatives
\be
R^{bcde}\nabla^2R_{bcde}=4R^{bcde}\nabla_b\nabla_dR_{ce}+2R^{bcde}R^f_bR_{fcde}+2R^{bcde}R^{a \ \ f}_{ \ bc \ }R_{afde}+4R^{bcde}R^{a\ \ f}_{ \ bd \ }R_{acfe}.
\ee
The first term can be written as (neglecting total derivatives),
\be
4R^{bcde}\nabla_b\nabla_dR_{ce}=-4\nabla_bR^{bcde}\nabla_dR_{ce}
=-4(\nabla_dR_{ce})^2-4R^{cd}\nabla^e\nabla_dR_{ce}\,,
\ee
and
\be
-4R^{cd}\nabla^e\nabla_dR_{ce}=(\nabla R)^2-4R^{cd}R^{e \ \ f}_{ \ dc \ }R_{ef}-4R^c_dR^d_fR^f_c.
\ee

\section{ Holographic stress tensor involving $\nabla R$ terms}\label{holstress}
Here we consider an extended analysis of \cite{FM} to derive the holographic stress tensor including the $\nabla\dots\nabla R$ terms in the action. The most general analysis is deferred for future work although from the following analysis it will be clear that the most general case will also work out in an analogous way. We consider the most general term involving two $\nabla$s in the action. Such terms after background field expansion are schematically given by
\be
S=\int d^{d+1}x \sqrt{g} Z \nabla\nabla (\D R)^n,
\ee
where $Z$ contains all the relevant tensor structures. Note that the Wald functional obtained from such a terms will be of the form
\be
E_R^{abcd}=\dots+Z^{ef\dots}\nabla_e\nabla_f (\D R)^{n-1}+\dots.
\ee
For $n>2$, these terms vanish since when we put the background AdS, $\D R_{abcd}$ vanishes in the variation of $E_R^{abcd}$. So the only terms at the two $\nabla$s order relevant for the calculation of the holographic stress tensor are schematically given by $\nabla \D R\nabla \D R$. These terms in the action are:
\be
S_{\nabla R}=\int d^{d+1}x \sqrt{g}Z^{efabcdmnrs}\D R_{mnrs}\nabla_e\nabla_f\D R_{abcd}\,,
\ee
where as before $Z^{efabcdmnrs}$ contains all possible tensor structures.\\
\par We now focus on the derivation of the holographic stress tensor for the action including \eqref{nablar}. The Wald functional corresponding to this term is given by
\be
E_{\nabla R}^{abcd}=d_1 g^{\la ab}g^{cd\ra}\nabla^2 \D R+d_2 \nabla^2 \D R^{\la ab}g^{cd\ra}+d_3 \nabla^2\D R^{abcd}\,,
\ee
and evaluated on the AdS, $E_{\nabla R}^{abcd}=0$, while the linearized variation of the wald function is given by
\be
\d E_{\nabla R}^{abcd}=\d ({\mathcal{Z}}^{efabcdmnrs}\nabla_e\nabla_f\D R_{mnrs}) ,
\ee
where the structure of ${\mathcal{Z}}$ for the contributing terms is given by
\be
{\mathcal{Z}}^{efabcdmnrs}=g^{ef}(d_1 g^{ac}g^{bd}g^{mr}g^{ns}+d_2 g^{ac}g^{mr}g^{bn}g^{ds}+d_3 g^{am}g^{bn}g^{cr}g^{ds}).
\ee
All indices are raised or lowered with respect to the background AdS metric $g_{ab}$. Thus combined with the original expressions in \cite{FM} for $E_R^{(1)abcd}=E_R^{abcd}+E_{\nabla R}^{abcd}=E_R^{abcd} \ \text{and} \ \d E_R^{(1)abcd}$ is given by,
\be
\d E_R^{(1)abcd}=-c_2 h g^{\la ab}g^{cd\ra}-c_3 h^{\la ab}g^{cd\ra}+c_4 {\mathcal{R}} g^{\la ab}g^{cd\ra}+c_5 {\mathcal{R}}^{\la ab}g^{cd\ra}+c_6 {\mathcal{R}}^{abcd}+\d E_{\nabla R}^{abcd}\,.
\ee

\subsection{\underline{d=4}}
The coefficients \eqref{ew} for $d=4$ are given by
\begin{align}
\begin{split}
A&=-\frac{c_3}{4}-\frac{3c_5}{4}-5c_6+64d_3, \ B=-\frac{c_2}{2}-2c_4-c_5-24d_2+8d_3,\\
C&=-c_2-4c_4+c_5+64d_3,\ D=-c_3-3c_5+4c_6-128d_3.
\end{split}
\end{align}
Thus the coefficients $A_1 \ A_2$ take the form
\begin{align}
\begin{split}
A_1&=-24(c_6-16d_3)R^2+\frac{1}{2}(c_1-c_3-3c_5-68c_6+1024d_3)(\frac{r^2}{3}-R^2),\\
A_2&=(\frac{c_1}{2}-c_2-4c_4-2c_5-48d_2+16d_3)R^2+\frac{r^2}{6}(c_1-c_3+15c_5+4c_6+288d_2+160d_3).
\end{split}
\end{align}
We can use the tracelessness condition of $h_{\m\n}$ \textsl{viz.} $h^{(d)\m}_{\m}=0$ to eliminate $A_2$ and thus integrate over $A_1$ to get
\be
\d S_B^{wald}=\frac{8\pi \Omega_2 \tilde{L} R^4}{15}(c_1+4c_6-64d_3),
\ee
where $\Omega_2$ is the volume of the unit $S_2$ and finally using \eqref{Ttt}, we have
\be
\d T_{tt}^{grav}=4\tilde{L}[c_1+4(c_6-16d_3)]\,.
\ee
 
\subsection{\underline{d=6}}
The corresponding coefficients in \eqref{ew} for $d=6$ after putting $\D=d$ are given as
\begin{align}
\begin{split}
A&=\frac{1}{4}(-c_3-5c_5-52c_6+1152d_3),\ B=-\frac{c_2}{2}-3c_4-\frac{11}{4}c_5-60d_2+12d_3,\\
C&=-c_2-6c_4+2c_5-120d_2,\ D=-c_3-5c_5+8c_6-288d_3.
\end{split}
\end{align}
Putting these in the integral we have
\begin{align}
\begin{split}
A_1&=\frac{1}{2}(c_1-c_3-5c_5-292c_6+6912d_3)(\frac{r^2}{5}-R^2)-120(c_6-24d_3)R^2,\\
A_2&=\frac{1}{2}(c_1-2c_2-12c_4-11c_5-240d_2+48d_3)R^2+\frac{1}{2}(c_1-c_3+70c_5+8c_6-528d_3)\frac{r^2}{5}.
\end{split}
\end{align}
Again by using the tracelessness argument we can set $h=0$ and integrating and finally using \eqref{Ttt}, we get,
\be
\d T_{tt}^{grav}=\frac{35}{2\pi\Omega_4}\lim_{R\rightarrow0}(\frac{1}{R^6}\d S_B^{wald})=6 \tilde{L}^3[c_1+8(c_6-24d_3)].
\ee

\section{$\la T_{\m\n}(x)T_{\r\s}(0)\ra$ in even dimensions}\label{A}
The B-type anomaly coefficients appearing in the expression for the holographic stress tensor in even dimensions are precisely the coefficients of the stress tensor two point functions from the field theory perspective. The 2d and 4d cases were worked out in \cite{erdmenger}. We will extend this result to 6d in what follows. Before that we will review the 2d and 4d results.

The starting point of the derivation is the renormalization group equation in \cite{osborn}, \cite{erdmenger} which takes on the form in general $d$ dimensions as
\be
(\m\pd_{\m}+2\int d^d x g^{\m\n}\frac{\d}{\d g^{\m\n}}) W=0.
\ee
We know that
\be
\int d^d x  g^{\m\n}\frac{\d}{\d g^{\m\n}}W=\int d^d x g^{\m\n}\la T_{\m\n}\ra=\int d^d x A_{anomaly},
\ee
which gives us
\be
\m \pd_{\m}W=-2\int d^d x A_{anomaly}.
\ee
We now functionally differentiate the LHS w.r.t $g^{\m\n}$ twice to get
\be
\m\pd_{\m}\la T_{ab}(x)T_{cd}(0)\ra=-2\int d^dx \frac{\d^2 A_{anomaly}}{\d g^{ab}\d g^{cd}}.
\ee
From the general conformal properties of the 2 point functions the RHS now takes the form
\be\label{2pt}
\m\pd_{\m}\la T_{ab}(x)T_{cd}(0)\ra=\frac{{\mathcal{C}}_T}{4(d-2)^2 d(d+1)} \D^T_{abcd}\m\pd_{\m}\frac{1}{x^{2d-4}},
\ee
where the tensor $\D^T_{abcd}$ now takes the form
\be
\D^T_{abcd}=\frac{1}{2}(S_{ac}S_{bd}+S_{ad}S_{bc})-\frac{1}{d-1}S_{ab}S_{cd}, \ \ \D^T_{aacd}=0,
\ee
where $S_{ab}=\pd_a\pd_b-\d_{ab}\pd^2$.
In general $x^{-2d+4}$ is singular function. We need to regularize the function in what follows.\\

\subsection{\underline{d=2}}
We consider the anomaly in $d=2$ which is given by $E_2=\frac{c}{4}R, \ I_2=0$. The RG equation is given by
\be
\m\pd_\m W+\int d^2 x \la T^i_i\ra=0,
\ee
and the 2 pt function is given by
\be
\m\pd_\m \la T_{ab}(x)T_{cd}(0)\ra=\frac{c}{24\pi}\int d^2 x \frac{\d^2 R}{\d g^{ab}\d g^{cd}}.
\ee
From the second order variation of $R$, $\d^2 R=h\pd^2 h-h \pd_e\pd_f h^{ef}$ we get,
\be
\frac{\d^2 R}{\d g^{ab}\d g^{cd}}=[(g_{ab}\pd_c\pd_d+g_{cd}\pd_a\pd_b)-g_{ab}g_{cd}\pd^2]\d^2(x).
\ee
Converting into the momentum space we can see that
\be
\m\pd_\m \la T_{ab}(p)T_{cd}(0)\ra=\frac{c}{24\pi}[(g_{ab}p_cp_d+g_{cd}p_ap_b)-g_{ab}g_{cd}p^2]\,,
\ee
using which we see that $\mathcal{C}_T$ and $c$ are proportional to one another.

\subsection{\underline{d=4}}
In 4d there are two anomalies given by
\begin{align}
\begin{split}
E_4&=R^{abcd}R_{abcd}-4R^{ab}R_{ab}+R^2\,,\\
I_4&=E_4+2(R^{ab}R_{ab}-\frac{1}{3}R^2).
\end{split}
\end{align}
The contribution from the $E_4$ term in 4$d$ is given by the integral of
\be
\int d^d x {\mathcal{A}}^E_{\r\s,\a\b}(x-y,x-z),
\ee
where the term ${\mathcal{A}}^E_{\r\s,\a\b}(x-y,x-z)$ is given by
\be\label{eqGB}
{\mathcal{A}}^E_{\r\s,\a\b}(x-y,x-z)=-(\e_{\s\a\g\k}\e_{\r\b\d\l}\pd_{\k}\pd_{\l}(\pd_{\g}\d^d(x-y)\pd_{\d}\d^d(x-z))+\e_{\r\a\g\k}\e_{\s\b\d\l}\pd_{\k}\pd_{\l}(\pd_{\g}\d^d(x-y)\pd_{\d}\d^d(x-z))).
\ee
To compute the integral we first convert the $\d^d(x-y)$ into momentum space and carry out the differentiations as
\be
\pd_\d \d^d(x-z)\pd_\g \d^d(x-y)=\pd_\g(\int e^{i p(x-y)}d^d p) \ \pd_\d(\int e^{i q(x-z)}d^d q)=-p_{\g}q_{\d}\int e^{i(p+q)x-ipy-iqz}d^d p \ d^d q.
\ee
Acting $\pd_\k\pd_\l$ on this, we get
\be
\pd_\k\pd_\l(-p_{\g}q_{\d}\int e^{i(p+q)x-ipy-iqz}d^d p \ d^d q)=p_{\g}q_\d(p+q)_\l(p+q)_\k \int e^{i(p+q)x-ipy-iqz}d^d p \ d^d q.
\ee
Thus the first term on the lhs in the above integral (\ref{eqGB}) becomes 
\be
\e_{\s\a\g\k}\e_{\r\b\d\l}p_{\g}q_{\d}(p+q)_{\l}(p+q)_{\k}\int e^{i(p+q)x}d^d x\int e^{-ipy-iqz}d^d p  \ d^d q\,,
\ee
which becomes after substituting the delta function from the first integral as
\be
\e_{\s\a\g\k}\e_{\r\b\d\l}\int p_{\g}q_{\d}(p+q)_{\l}(p+q)_{\k}\d^d(p+q) e^{-ipy-iqz}d^d p  \ d^d q.
\ee
Thus this integral vanishes on its own. Similarly it can be shown that the second part of the integral also vanishes by itself. Thus there is no contribution from the $E_4$ term to the anomaly. The only contribution to the anomaly comes from the term $R^{ab}R_{ab}-\frac{1}{3}R^2$ term in the Weyl anomaly. Thus
\be
\m\pd_\m \la T_{ab}(x)T_{cd}(0)\ra=\frac{c}{16\pi^2}\int d^4 x \frac{\d^2}{\d g^{ab}g^{cd}}[2(R^{mn}R_{mn}-\frac{1}{3}R^2)].
\ee
The last term on the rhs gives
\be
\frac{\d^2 R^2}{\d g^{ab}\g ^{cd}}=2\frac{\d R}{\d g^{ab}}\frac{\d R}{\d g^{cd}}.
\ee
After linearization of the scalar and functionally differentiating w.r.t $g^{ab}$ we have
\be
\frac{\d R}{\d g^{ab}}=(\pd_a\pd_b-g_{ab}\pd^2)\d^4(x)=S_{ab}\d^4(x),
\ee
where we define $S_{ab}=\pd_a\pd_b-g_{ab}\pd^2$.
Thus the last term becomes after some integration by parts
\be
\frac{\d^2 R^2}{\d g^{ab}\d g^{cd}}=S_{ab}S_{cd}\d^{4}(x).
\ee
The first term on the rhs becomes after integration by parts as
\be
\frac{\d R^{mn}}{\d g^{ab}}\frac{\d R_{mn}}{\d g^{cd}}=\frac{1}{2}(S_{ac}S_{bd}+S_{ad}S_{bc})\d^4(x).
\ee
Thus the total contribution from the Weyl anomaly is given by
\be
\m\pd_\m \la T_{ab}(x)T_{cd}(0)\ra=-4\b\D^{T}_{abcd}\d^4(x)\,,
\ee
where we define $\D^T_{abcd}=\frac{1}{2}(S_{ac}S_{bd}+S_{ad}S_{bc})-\frac{1}{3}S_{ab}S_{cd}$.

Thus in 4$d$ we have using $\b=-c/16\pi^2$ from \cite{osborn}
\be
\m\pd_{\m}\la T_{ab}(x)T_{cd}(0)\ra=\frac{c}{4\pi^2}\D^T_{abcd}\d^4(x).
\ee
Comparing with (\ref{2pt}) we have 
\be\label{ct}
\frac{{\mathcal{C}}_T}{4(d-2)^2d(d+1)}\m\pd_{\m}\frac{1}{x^4}=\frac{c}{4\pi^2}\d^4(x).
\ee
In 4d the regularized $1/x^4$ can be expressed as
\be
{\mathcal{R}}\frac{1}{x^4}=-\frac{1}{4}\pd^2\frac{1}{x^2}(\log \m^2x^2) \ \Rightarrow \ \m\pd_{\m}{\mathcal{R}}\frac{1}{x^4}=2\pi^2 \d^4(x).
\ee
Putting this in (\ref{ct}) we have 
\be
c=\frac{\pi^4}{40}{\mathcal{C}}_T.
\ee

\subsection{\underline{d=6}}
In 6d it is rather easy to see why only the $B_3$ coefficient gets picked up by the 2 pt functions. If we look at the structures of the anomalies then only $I_3$ has a structure of the form
\be
I_3\sim C^{abcd}\pd^2 C_{abcd}.
\ee
This makes $I_3$ to start at the order $O(h^2)$ and contributes in the 2 pt function. While all the other anomalies start at $O(h^3)$ and thus do not contribute.

In 6d the only contribution to the two point function comes from the term $I_3\sim C^{abcd}\pd^2 C_{abcd}$, since the other anomalies start at $O(h^3)$.
Thus from  (\ref{2pt}) we have 
\be
\m\pd_{\m}\la T_{ab}(x)T_{cd}(0)\ra=6B_3 \D^T_{abcd}\pd^2\d^6(x)=\D^T_{abcd}\frac{{\mathcal{C}}_T}{2^7\times3\times 7}\m\pd_\m \frac{1}{x^8}.
\ee
In 6d we regularize as
\be
{\mathcal{R}}\frac{1}{x^8}=-\frac{1}{96}\pd^4\frac{1}{x^4}(\log \m^2x^2)\ \Rightarrow \ \m\pd_{\m}{\mathcal{R}}\frac{1}{x^8}=-\frac{1}{48}\pd^4\frac{1}{x^4}.
\ee
The term on the RHS for 6d can be reduced to $\pd^4\frac{1}{x^4}=-4\pi^3 \pd^2 \d^6(x)$ and hence the RHS becomes 
\be
\m\pd_{\m}{\mathcal{R}}\frac{1}{x^8}=-\frac{1}{48}\pd^4\frac{1}{x^4}=\frac{\pi^3}{12} \pd^2 \d^6(x).
\ee
Thus in 6d we have
\be
\m\pd_{\m}\la T_{ab}(x)T_{cd}(0)\ra=6B_3\D^T_{abcd}\pd^2\d^6(x)=\frac{\pi^3}{7\times3^2\times2^{9}}\D^T_{abcd}{\mathcal{C}}_T\pd^2\d^{6}(x).
\ee
Hence we have
\be 
B_3=\frac{\pi^3}{64\cdot 7!}\frac{5}{3}{\mathcal{C}}_T.
\ee

\section{ $\eta/s$ for general $R^2$ theories}\label{etas}
We will calculate the ratio of the shear viscosity to the entropy density for four derivative theory of gravity in $d=4$ where $d$ is the boundary dimension. We want to express the ratio in terms of the field theory variables as $t_2\ \text{etc}.$ This analysis can be extended for general higher derivative theories of gravity in arbitrary dimensions. To proceed we will follow the analysis of \cite{miguel} where the horizon is first constructed perturbatively and then the pole method was used to extract the shear viscosity. We first consider the metric as
\be
ds^2=\frac{\tilde{L}^2}{4f(z)}\frac{dz^2}{(1-z)^2}+\frac{r_0^2}{\tilde{L}^2(1-z)}[-\frac{f(z)}{\fin}dt^2+(dx_1+\phi(t) dx_2)^2+dx_2^2+dx_3^2].
\ee
where $\phi(t)=e^{-i\omega t}$ is the fluctuation and 
\be\label{ff}
f(z)=2z+f_2 z^2+f_3 z^3+\dots.
\ee
We consider the general $R^2$ action given by
\be\label{genR2}
S=\int d^5x\sqrt{g}[R+\frac{12}{L^2}+\frac{L^4}{2}(\l_1 R^{abcd}R_{abcd}+\l_2 R^{ab}R_{ab}+\l_3 R^2)].
\ee
To obtain the coefficients $f_2, \ f_3$ we plug in \eqref{ff} into the equations of motion for the action \eqref{genR2} and solve perturbatively near the horizon. The solution for $\fin$ taking $L=\tilde{L}\sqrt{\fin}$ given by,
\be
1-\fin+\frac{1}{3}(\l_1+2 \l_2+10\l_3)\fin^2=0.
\ee
The expression for $c_1$ is given by
\be
c_1=\frac{1}{16\pi}[1-2\fin(\l_1+2\l_2+10\l_3].
\ee
We also express $c_6=\frac{\l_1}{16\pi}\fin$. The shear viscosity and entropy density in terms of these couplings are given by
\begin{align}
\begin{split}
\eta&={\mathcal{C}}_1[8\l_1^2+\l_2+4\l_3-20\l_2\l_3-64\l_3^2+12\l_1(\l_2+2\l_3)+\sqrt{{\mathcal{F}}}],\\
s&={\mathcal{C}}_2[16\l_1^2+\l_2+4\l_3+20\l_2\l_3+48\l_1\l_3-20\l_2\l_3-64\l_3^2+\sqrt{{\mathcal{F}}}],
\end{split}
\end{align}
where the normalizations are ${\mathcal{C}}_2=\frac{2\pi r_0^3\fin^{3/2}}{\ell_p^3}\ \text{and} \ {\mathcal{C}}_1=\frac{r_0^3\fin^{3/2}}{2\ell_p^3}$ and we have set $\tilde{L}=1$.
\begin{align}
\begin{split}
{\mathcal{F}}&=(2\l_1+\l_2+2\l_3)[(2\l_1+\l_2+2\l_3)(1-12\l_1-16\l_2-52\l_3)^2\\
&-16(\l_1+\l_2+2\l_3)(22\l_1^2-\l_2(1-12\l_2)-2\l_3+62\l_2\l_3+70\l_3^2
-2\l_1(1-19\l_2-58\l_3))]\,.
\end{split}
\end{align}
The corresponding expression for $t_2$ is $d=4$ is given by
\be
t_2=\frac{24 c_6}{c_1+4c_6}=\frac{24\l_1\fin}{1+2(\l_1-2\l_2-10\l_3)\fin}.
\ee
Note that in the limit $\l_1,\l_2,\l_3\rightarrow0$ we retrieve the result
\be
\frac{\eta}{s}=\frac{1}{4\pi}.
\ee
Note also that it is possible to make $\eta/s$ arbitrarily small by tuning the values of $\l$s'. For example for $\l_1=0.31517, \ \l_2=\l_3=-1$, we have
\be
\frac{\eta}{s}=\frac{1.1\times 10^{-5}}{4\pi}.
\ee
Here the constraints arising from $\la \epsilon \ra >0$ are satisfied.


\begin{thebibliography}{99}

\bibitem{revs} 
  D.~T.~Son and A.~O.~Starinets,
  ``Viscosity, Black Holes, and Quantum Field Theory,''
  Ann.\ Rev.\ Nucl.\ Part.\ Sci.\  {\bf 57}, 95 (2007)
  [arXiv:0704.0240 [hep-th]].\\
S.~A.~Hartnoll,
  ``Lectures on holographic methods for condensed matter physics,''
  Class.\ Quant.\ Grav.\  {\bf 26}, 224002 (2009)
  [arXiv:0903.3246 [hep-th]].\\
J.~Casalderrey-Solana, H.~Liu, D.~Mateos, K.~Rajagopal and U.~A.~Wiedemann,
  ``Gauge/String Duality, Hot QCD and Heavy Ion Collisions,''
  arXiv:1101.0618 [hep-th].

\bibitem{skenderis}
 S.~de Haro, S.~N.~Solodukhin and K.~Skenderis,
``Holographic reconstruction of space-time and renormalization in the AdS / CFT correspondence,''
  Commun.\ Math.\ Phys.\  {\bf 217}, 595 (2001)
  [hep-th/0002230]
K.~Skenderis,
``Lecture notes on holographic renormalization,''
Class.Quant.Grav.19:5849-5876,2002
[arXiv: hep-th/0209067]

\bibitem{HT}
 O.~Hohm and E.~Tonni,
  ``A boundary stress tensor for higher-derivative gravity in AdS and Lifshitz backgrounds,''
  JHEP {\bf 1004}, 093 (2010)
  [arXiv:1001.3598 [hep-th]]

\bibitem{others}
J.~Smolic and M.~Taylor,
  ``Higher derivative effects for 4d AdS gravity,''
  JHEP {\bf 1306}, 096 (2013)
  [arXiv:1301.5205 [hep-th]].\\
N.~Johansson, A.~Naseh and T.~Zojer,
  ``Holographic two-point functions for 4d log-gravity,''
  JHEP {\bf 1209}, 114 (2012)
  [arXiv:1205.5804 [hep-th]].\\
S.~-J.~Hyun, W.~-J.~Jang, J.~-H.~Jeong and S.~-H.~Yi,
  ``Noncritical Einstein-Weyl Gravity and the AdS/CFT Correspondence,''
  JHEP {\bf 1201}, 054 (2012)
  [arXiv:1111.1175 [hep-th]].\\
G.~Giribet, A.~Goya and M.~Leston,
  ``Boundary stress tensor and asymptotically AdS3 non-Einstein spaces at the chiral point,''
  Phys.\ Rev.\ D {\bf 84}, 066003 (2011)
  [arXiv:1108.0400 [hep-th]].\\
Y.~Kwon, S.~Nam, J.~-D.~Park and S.~-H.~Yi,
  ``Holographic Renormalization and Stress Tensors in New Massive Gravity,''
  JHEP {\bf 1111}, 029 (2011)
  [arXiv:1106.4609 [hep-th]].\\
E.~A.~Bergshoeff, O.~Hohm, J.~Rosseel and P.~K.~Townsend,
  ``Modes of Log Gravity,''
  Phys.\ Rev.\ D {\bf 83}, 104038 (2011)
  [arXiv:1102.4091 [hep-th]].\\
G.~Giribet and M.~Leston,
  ``Boundary stress tensor and counterterms for weakened AdS3 asymptotic in New Massive Gravity,''
  JHEP {\bf 1009}, 070 (2010)
  [arXiv:1006.3349 [hep-th]].
M.~Alishahiha and A.~Naseh,
  ``Holographic renormalization of new massive gravity,''
  Phys.\ Rev.\ D {\bf 82}, 104043 (2010)
  [arXiv:1005.1544 [hep-th]]
D.~Allahbakhshi, M.~Alishahiha and A.~Naseh,
  ``Entanglement Thermodynamics,''
  JHEP {\bf 1308}, 102 (2013)
  [arXiv:1305.2728 [hep-th]]\\
S.~Cremonini, J.~T.~Liu and P.~Szepietowski,
  ``Higher Derivative Corrections to R-charged Black Holes: Boundary Counterterms and the Mass-Charge Relation,''
  JHEP {\bf 1003}, 042 (2010)
  [arXiv:0910.5159 [hep-th]]


\bibitem{FM}
T.~Faulkner, M.~Guica, T.~Hartman, R.~C.~Myers and M.~Van Raamsdonk,
  ``Gravitation from Entanglement in Holographic CFTs,''
  JHEP {\bf 1403}, 051 (2014)
  [arXiv:1312.7856 [hep-th]]

\bibitem{tadashi}
 J.~Bhattacharya, M.~Nozaki, T.~Takayanagi and T.~Ugajin,
``Thermodynamical Property of Entanglement Entropy for Excited States,''
 Phys.\ Rev.\ Lett.\  {\bf 110}, no. 9, 091602 (2013)
 [arXiv:1212.1164]\\
D.~D.~Blanco, H.~Casini, L.~-Y.~Hung and R.~C.~Myers,
  ``Relative Entropy and Holography,''
  JHEP {\bf 1308}, 060 (2013)
  [arXiv:1305.3182 [hep-th]].

\bibitem{IW}
 V.~Iyer and R.~M.~Wald,
  ``Some properties of Noether charge and a proposal for dynamical black hole entropy,''
  Phys.\ Rev.\ D {\bf 50}, 846 (1994)
  [gr-qc/9403028]

\bibitem{chinese}
R.~-X.~Miao,
  ``A Note on Holographic Weyl Anomaly and Entanglement Entropy,''
  Class.\ Quant.\ Grav.\  {\bf 31}, 065009 (2014)
  [arXiv:1309.0211 [hep-th]]

\bibitem{holGB}
A.~Buchel, J.~Escobedo, R.~C.~Myers, M.~F.~Paulos, A.~Sinha and M.~Smolkin,
  ``Holographic GB gravity in arbitrary dimensions,''
  JHEP {\bf 1003}, 111 (2010)
  [arXiv:0911.4257 [hep-th]]

\bibitem{HM}
D.~M.~Hofman and J.~Maldacena,
  ``Conformal collider physics: Energy and charge correlations,''
  JHEP {\bf 0805}, 012 (2008)
  [arXiv:0803.1467 [hep-th]]

\bibitem{hofman}
 D.~M.~Hofman,
  ``Higher Derivative Gravity, Causality and Positivity of Energy in a UV complete QFT,''
  Nucl.\ Phys.\ B {\bf 823}, 174 (2009)
  [arXiv:0907.1625 [hep-th]].

\bibitem{quasitop}
 R.~C.~Myers, M.~F.~Paulos and A.~Sinha,
  ``Holographic studies of quasi-topological gravity,''
  JHEP {\bf 1008}, 035 (2010)
  [arXiv:1004.2055 [hep-th]]
 D.~M.~Hofman,
  ``Higher Derivative Gravity, Causality and Positivity of Energy in a UV complete QFT,''
  Nucl.\ Phys.\ B {\bf 823}, 174 (2009)
  [arXiv:0907.1625 [hep-th]]

\bibitem{osborn}
 H.~Osborn and A.~C.~Petkou,
  ``Implications of conformal invariance in field theories for general dimensions,''
  Annals Phys.\  {\bf 231}, 311 (1994)
  [hep-th/9307010]

\bibitem{erdmenger}
J.~Erdmenger and H.~Osborn,
 ``Conserved currents and the energy momentum tensor in conformally invariant theories for general dimensions,''
  Nucl.\ Phys.\ B {\bf 483}, 431 (1997)
  [hep-th/9605009]

\bibitem{RM}
 R.~C.~Myers and B.~Robinson,
  ``Black Holes in Quasi-topological Gravity,''
  JHEP {\bf 1008}, 067 (2010)
  [arXiv:1003.5357 [gr-qc]]

\bibitem{etabs}
 A.~Buchel, J.~T.~Liu and A.~O.~Starinets,
  ``Coupling constant dependence of the shear viscosity in N=4 supersymmetric Yang-Mills theory,''
  Nucl.\ Phys.\ B {\bf 707}, 56 (2005)
  [hep-th/0406264].\\
 R.~C.~Myers, M.~F.~Paulos and A.~Sinha,
  ``Quantum corrections to eta/s,''
  Phys.\ Rev.\ D {\bf 79}, 041901 (2009)
  [arXiv:0806.2156 [hep-th]].\\
A.~Buchel, R.~C.~Myers, M.~F.~Paulos and A.~Sinha,
  ``Universal holographic hydrodynamics at finite coupling,''
  Phys.\ Lett.\ B {\bf 669}, 364 (2008)
  [arXiv:0808.1837 [hep-th]].\\
A.~Buchel, R.~C.~Myers and A.~Sinha,
  ``Beyond eta/s = 1/4 pi,''
  JHEP {\bf 0903}, 084 (2009)
  [arXiv:0812.2521 [hep-th]].\\
A.~Sinha and R.~C.~Myers,
  ``The Viscosity bound in string theory,''
  Nucl.\ Phys.\ A {\bf 830}, 295C (2009)
  [arXiv:0907.4798 [hep-th]].\\
N.~Banerjee and S.~Dutta,
  ``Shear Viscosity to Entropy Density Ratio in Six Derivative Gravity,''
  JHEP {\bf 0907}, 024 (2009)
  [arXiv:0903.3925 [hep-th]].\\
X.~O.~Camanho, J.~D.~Edelstein and M.~F.~Paulos,
  ``Lovelock theories, holography and the fate of the viscosity bound,''
  JHEP {\bf 1105}, 127 (2011)
  [arXiv:1010.1682 [hep-th]].\\
R.~Brustein and A.~J.~M.~Medved,
  ``Non-perturbative unitarity constraints on the ratio of shear viscosity to entropy density in UV complete theories with a gravity dual,''
  Phys.\ Rev.\ D {\bf 84}, 126005 (2011)
  [arXiv:1108.5347 [hep-th]]
A.~Buchel, J.~Escobedo, R.~C.~Myers, M.~F.~Paulos, A.~Sinha and M.~Smolkin,
  ``Holographic GB gravity in arbitrary dimensions,''
  JHEP {\bf 1003}, 111 (2010)
  [arXiv:0911.4257 [hep-th]]\\
R.~C.~Myers, M.~F.~Paulos and A.~Sinha,
  ``Holographic Hydrodynamics with a Chemical Potential,''
  JHEP {\bf 0906}, 006 (2009)
  [arXiv:0903.2834 [hep-th]].\\
S.~Cremonini, K.~Hanaki, J.~T.~Liu and P.~Szepietowski,
  ``Higher derivative effects on eta/s at finite chemical potential,''
  Phys.\ Rev.\ D {\bf 80}, 025002 (2009)
  [arXiv:0903.3244 [hep-th]]

\bibitem{vis}
 M.~Brigante, H.~Liu, R.~C.~Myers, S.~Shenker and S.~Yaida,
  ``Viscosity Bound Violation in Higher Derivative Gravity,''
  Phys.\ Rev.\ D {\bf 77}, 126006 (2008)
  [arXiv:0712.0805 [hep-th]].\\
M.~Brigante, H.~Liu, R.~C.~Myers, S.~Shenker and S.~Yaida,
  ``The Viscosity Bound and Causality Violation,''
  Phys.\ Rev.\ Lett.\  {\bf 100}, 191601 (2008)
  [arXiv:0802.3318 [hep-th]]



\bibitem{KSS}
P.~Kovtun, D.~T.~Son and A.~O.~Starinets,
  ``Viscosity in strongly interacting quantum field theories from black hole physics,''
  Phys.\ Rev.\ Lett.\  {\bf 94}, 111601 (2005)
  [hep-th/0405231]\\
P.~Kovtun, D.~T.~Son and A.~O.~Starinets,
  ``Holography and hydrodynamics: Diffusion on stretched horizons,''
  JHEP {\bf 0310}, 064 (2003)
  [hep-th/0309213]\\
G.~Policastro, D.~T.~Son and A.~O.~Starinets,
  ``The Shear viscosity of strongly coupled N=4 supersymmetric Yang-Mills plasma,''
  Phys.\ Rev.\ Lett.\  {\bf 87}, 081601 (2001)
  [hep-th/0104066]




\bibitem{sinha}L.~-Y.~Hung, R.~C.~Myers and M.~Smolkin,
  ``On Holographic Entanglement Entropy and Higher Curvature Gravity,''
  JHEP {\bf 1104}, 025 (2011)
  [arXiv:1101.5813 [hep-th]].\\
J.~de Boer, M.~Kulaxizi and A.~Parnachev,
  ``Holographic Entanglement Entropy in Lovelock Gravities,''
  JHEP {\bf 1107}, 109 (2011)
  [arXiv:1101.5781 [hep-th]].\\
 A.~Bhattacharyya, A.~Kaviraj and A.~Sinha,
  ``Entanglement entropy in higher derivative holography,''
  JHEP {\bf 1308}, 012 (2013)
  [arXiv:1305.6694 [hep-th]].\\
D.~V.~Fursaev, A.~Patrushev and S.~N.~Solodukhin,
``Distributional Geometry of Squashed Cones.''
Phys.Rev.D88(2013)4, 044054
[arXiv: 1306.4000[hep-th]].\\
A.~Bhattacharyya, M.~Sharma and A.~Sinha,
  ``On generalized gravitational entropy, squashed cones and holography,''
  JHEP {\bf 1401}, 021 (2014)
  [arXiv:1308.5748 [hep-th]].\\
 X.~Dong,
  ``Holographic Entanglement Entropy for General Higher Derivative Gravity,''
  JHEP {\bf 1401}, 044 (2014)
  [arXiv:1310.5713 [hep-th], arXiv:1310.5713].\\
 J.~Camps,
  ``Generalized entropy and higher derivative Gravity,''
  JHEP {\bf 1403}, 070 (2014)
  [arXiv:1310.6659 [hep-th]].\\
A.~Bhattacharyya and M.~Sharma,
  ``On entanglement entropy functionals in higher derivative gravity theories,''
  arXiv:1405.3511 [hep-th].

\bibitem{weylanom} 
  M.~Henningson and K.~Skenderis,
  ``The Holographic Weyl anomaly,''
  JHEP {\bf 9807}, 023 (1998)
  [hep-th/9806087].

\bibitem{nojiri}
S.~'i.~Nojiri and S.~D.~Odintsov,
  ``On the conformal anomaly from higher derivative gravity in AdS / CFT correspondence,''
  Int.\ J.\ Mod.\ Phys.\ A {\bf 15}, 413 (2000)
  [hep-th/9903033]

\bibitem{SSS}
 K.~Sen, A.~Sinha and N.~V.~Suryanarayana,
  ``Counterterms, critical gravity and holography,''
  Phys.\ Rev.\ D {\bf 85}, 124017 (2012)
  [arXiv:1201.1288 [hep-th]]

\bibitem{imbimbo} 
  C.~Imbimbo, A.~Schwimmer, S.~Theisen and S.~Yankielowicz,
  ``Diffeomorphisms and holographic anomalies,''
  Class.\ Quant.\ Grav.\  {\bf 17}, 1129 (2000)
  [hep-th/9910267].

\bibitem{deserschwimmer} 
  S.~Deser and A.~Schwimmer,
  ``Geometric classification of conformal anomalies in arbitrary dimensions,''
  Phys.\ Lett.\ B {\bf 309}, 279 (1993)
  [hep-th/9302047].

\bibitem{6danom} 
  F.~Bastianelli, S.~Frolov and A.~A.~Tseytlin,
  ``Conformal anomaly of (2,0) tensor multiplet in six-dimensions and AdS / CFT correspondence,''
  JHEP {\bf 0002}, 013 (2000)
  [hep-th/0001041].

\bibitem{deser}
S.~Deser, H.~Liu, H.~Lu, C.~N.~Pope, T.~C.~Sisman and B.~Tekin,
  ``Critical Points of D-Dimensional Extended Gravities,''
  Phys.\ Rev.\ D {\bf 83}, 061502 (2011)
  [arXiv:1101.4009 [hep-th]]

\bibitem{liutseytlin} 
  H.~Liu and A.~A.~Tseytlin,
  ``D = 4 superYang-Mills, D = 5 gauged supergravity, and D = 4 conformal supergravity,''
  Nucl.\ Phys.\ B {\bf 533}, 88 (1998)
  [hep-th/9804083].

\bibitem{edelstein}
X.~O.~Camanho, J.~D.~Edelstein and J.~M.~Sánchez De Santos,
  ``Lovelock theory and the AdS/CFT correspondence,''
  Gen.\ Rel.\ Grav.\  {\bf 46}, 1637 (2014)
  [arXiv:1309.6483 [hep-th]]\\
J.~D.~Edelstein,
  ``Lovelock theory, black holes and holography,''
  arXiv:1303.6213 [gr-qc]\\
X.~O.~Camanho and J.~D.~Edelstein,
  ``Causality in AdS/CFT and Lovelock theory,''
  JHEP {\bf 1006}, 099 (2010)
  [arXiv:0912.1944 [hep-th]]
 X.~O.~Camanho and J.~D.~Edelstein,
  ``Causality constraints in AdS/CFT from conformal collider physics and Gauss-Bonnet gravity,''
  JHEP {\bf 1004}, 007 (2010)
  [arXiv:0911.3160 [hep-th]]

\bibitem{horowitz}
G.~T.~Horowitz and N.~Itzhaki,
  ``Black holes, shock waves, and causality in the AdS / CFT correspondence,''
  JHEP {\bf 9902}, 010 (1999)
  [hep-th/9901012]

\bibitem{bpk}
J.~de Boer, M.~Kulaxizi and A.~Parnachev,
  ``Holographic Lovelock Gravities and Black Holes,''
  JHEP {\bf 1006}, 008 (2010)
  [arXiv:0912.1877 [hep-th]].\\
J.~de Boer, M.~Kulaxizi and A.~Parnachev,
  ``AdS(7)/CFT(6), Gauss-Bonnet Gravity, and Viscosity Bound,''
  JHEP {\bf 1003}, 087 (2010)
  [arXiv:0910.5347 [hep-th]]

\bibitem{miguel}
M.~F.~Paulos,
  ``Transport coefficients, membrane couplings and universality at extremality,''
  JHEP {\bf 1002}, 067 (2010)
  [arXiv:0910.4602 [hep-th]]

\bibitem{db}
N.~Banerjee and S.~Dutta,
  ``Near-Horizon Analysis of eta/s,''
  Nucl.\ Phys.\ B {\bf 845}, 165 (2011)
  [arXiv:0911.0557 [hep-th]]

\bibitem{athm}
J.~L.~Cardy,
  ``Is There a c Theorem in Four-Dimensions?,''
  Phys.\ Lett.\ B {\bf 215}, 749 (1988).\\
Z.~Komargodski and A.~Schwimmer,
  ``On Renormalization Group Flows in Four Dimensions,''
  JHEP {\bf 1112}, 099 (2011)
  [arXiv:1107.3987 [hep-th]].

\bibitem{cthm}
L.~Girardello, M.~Petrini, M.~Porrati and A.~Zaffaroni,
  ``Novel local CFT and exact results on perturbations of N=4 superYang Mills from AdS dynamics,''
  JHEP {\bf 9812}, 022 (1998)
  [hep-th/9810126].\\
D.~Z.~Freedman, S.~S.~Gubser, K.~Pilch and N.~P.~Warner,
  ``Renormalization group flows from holography supersymmetry and a c theorem,''
  Adv.\ Theor.\ Math.\ Phys.\  {\bf 3}, 363 (1999)
  [hep-th/9904017].\\
R.~C.~Myers and A.~Sinha,
  ``Seeing a c-theorem with holography,''
  Phys.\ Rev.\ D {\bf 82}, 046006 (2010)
  [arXiv:1006.1263 [hep-th]]\\
R.~C.~Myers and A.~Sinha,
  ``Holographic c-theorems in arbitrary dimensions,''
  JHEP {\bf 1101}, 125 (2011)
  [arXiv:1011.5819 [hep-th]].\\
A.~Sinha,
  ``On higher derivative gravity, $c$-theorems and cosmology,''
  Class.\ Quant.\ Grav.\  {\bf 28}, 085002 (2011)
  [arXiv:1008.4315 [hep-th]]\\
J.~T.~Liu, W.~Sabra and Z.~Zhao,
  ``Holographic c-theorems and higher derivative gravity,''
  Phys.\ Rev.\ D {\bf 85}, 126004 (2012)
  [arXiv:1012.3382 [hep-th]].\\
Y.~Nakayama,
  ``Does anomalous violation of null energy condition invalidate holographic c-theorem?,''
  Phys.\ Lett.\ B {\bf 720}, 265 (2013)
  [arXiv:1211.4628 [hep-th]].\\
A.~Bhattacharyya, L.~-Y.~Hung, K.~Sen and A.~Sinha,
  ``On c-theorems in arbitrary dimensions,''
  Phys.\ Rev.\ D {\bf 86}, 106006 (2012)
  [arXiv:1207.2333 [hep-th]].\\
M.~F.~Paulos,
  ``Holographic phase space: $c$-functions and black holes as renormalization group flows,''
  JHEP {\bf 1105}, 043 (2011)
  [arXiv:1101.5993 [hep-th]].

\bibitem{jaff} 
 C.~A.~Agon, M.~Headrick, D.~L.~Jafferis and S.~Kasko,
  ``Disk entanglement entropy for a Maxwell field,''
  Phys.\ Rev.\ D {\bf 89}, 025018 (2014)
  [arXiv:1310.4886 [hep-th]]

\bibitem{raju}
 S.~Raju,
  ``Four Point Functions of the Stress Tensor and Conserved Currents in AdS$_4$/CFT$_3$,''
  Phys.\ Rev.\ D {\bf 85}, 126008 (2012)
  [arXiv:1201.6452 [hep-th]]\\
A.~Dymarsky,
  ``On the four-point function of the stress-energy tensors in a CFT,''
  arXiv:1311.4546 [hep-th].

\bibitem{grumiller}
D.~Grumiller and O.~Hohm,
  ``AdS(3)/LCFT(2): Correlators in New Massive Gravity,''
  Phys.\ Lett.\ B {\bf 686}, 264 (2010)
  [arXiv:0911.4274 [hep-th]]\\
D.~Grumiller and I.~Sachs,
  ``AdS (3) / LCFT (2) ---Correlators in Cosmological Topologically Massive Gravity,''
  JHEP {\bf 1003}, 012 (2010)
  [arXiv:0910.5241 [hep-th]]\\
D.~Grumiller, M.~Irakleidou, I.~Lovrekovic and R.~McNees,
  ``Conformal gravity holography in four dimensions,''
  Phys.\ Rev.\ Lett.\  {\bf 112}, 111102 (2014)
  [arXiv:1310.0819 [hep-th]]

\bibitem{rel}
S.~Banerjee, A.~Bhattacharyya, A.~Kaviraj, K.~Sen and A.~Sinha,
  ``Constraining gravity using entanglement in AdS/CFT,''
  JHEP {\bf 1405}, 029 (2014)
  [arXiv:1401.5089 [hep-th]].












\end{thebibliography}
\end{document}